\documentclass[lettersize,journal]{IEEEtran}
\usepackage{booktabs}
\usepackage{threeparttable}
\usepackage{graphicx,color}
\usepackage[utf8]{inputenc}
\usepackage{hyperref}
\usepackage[T1]{fontenc}  
\usepackage{listings}
\usepackage{physics}
 \usepackage[detect-all]{siunitx}[=v2]
\usepackage{subcaption}
\usepackage{amsthm}
\usepackage{amsmath,amsfonts}
\usepackage{soul}
\usepackage{caption}
\usepackage{multirow,tabularx}
\usepackage{tcolorbox}
\usepackage{comment}
\usepackage{bm}
\usepackage{algorithm}
\usepackage{algpseudocode}
\usepackage{float}
\usepackage{tikz}
\usepackage{xcolor, colortbl}
\usepackage{cite}
\usepackage{amsmath,amssymb,amsfonts}
\usepackage{array}
\usepackage{graphicx}
\usepackage{stfloats}
\usepackage{textcomp}
\usepackage{changepage}
\DeclareMathOperator*{\argmin}{argmin} 
\DeclareMathOperator*{\argmax}{argmax} 
\usepackage{ifthen}
\usepackage[normalem]{ulem} 
\usepackage{xcolor}
\usepackage{amssymb}

\newboolean{showedits}
\setboolean{showedits}{true} 
\ifthenelse{\boolean{showedits}}
{
	\newcommand{\del}[1]{\textcolor{red}{\sout{#1}}} 
}{
	\newcommand{\del}[1]{} 
	
}

\newboolean{showcomments}
\setboolean{showcomments}{true} 
\newcommand{\id}[1]{$-$Id: scgPaper.tex 32478 2010-04-29 09:11:32Z oscar $-$}

\ifthenelse{\boolean{showcomments}}
{\newcommand{\nbc}[3]{
 {\colorbox{#3}{\bfseries\sffamily\scriptsize\textcolor{white}{#1}}}
 {\textcolor{#3}{\sf\small$\blacktriangleright$\textit{#2}$\blacktriangleleft$}}}
 }
{\newcommand{\nbc}[3]{}
 \renewcommand{\del}[1]{} 
 }

\definecolor{ibcolor}{rgb}{0.9,0.5,0}
\definecolor{cfcolor}{rgb}{0,0.5,0.9}
\definecolor{oldcolor}{rgb}{0.2,0.2,0.2}
\definecolor{tdcolor}{rgb}{1.0,0,0}
\definecolor{jycolor}{rgb}{0.7,0.3,0}
\definecolor{klcolor}{rgb}{0.13, 0.67, 0.8}

\newcommand\xy[1]{\nbc{XY}{#1}{ibcolor}}

\newcommand\todo[1]{\nbc{TODO}{#1}{tdcolor}}

\usepackage{xspace}

\newcommand\tool{UniAda\xspace}

\begin{document}

\title{UniAda: Universal Adaptive Multi-objective Adversarial Attack for End-to-End Autonomous Driving Systems}  

\author{Jingyu Zhang, Jacky Wai Keung, Yan Xiao, Yihan Liao, Yishu Li, and Xiaoxue Ma
\thanks{Manuscript received XX/XX/XX; revised XX/XX/XX; accepted XX/XX/XX. Date of publication XX/XX/XX.}
\thanks{This work is supported in part by the General Research Fund (GRF) of the Research Grants Council of Hong Kong, and the industry research funds of City University of Hong Kong (7005217, 9220097, 9220103, 9229029, 9229098, 9678149).
(\textit{Corresponding authors:} Yan Xiao.)}
\thanks{Jingyu Zhang, Jacky Wai Keung, Yihan Liao, Yishu Li and Xiaoxue Ma are with Department of Computer Science, City University of Hong Kong, Hong Kong SAR (e-mail: jzhang2297-c@my.cityu.edu.hk; Jacky.Keung@cityu.edu.hk; yihanliao4-c@my.cityu.edu.hk; yishuli5-c@my.cityu.edu.hk; xiaoxuema3-c@my.cityu.edu.hk}

\thanks{Yan Xiao is with School of Cyber Science and Technology, Shenzhen Campus of Sun Yat-sen University, Shenzhen, China (e-mail: xiaoyan.hhu@gmail.com).}

\thanks{Color versions of one or more figures in this article are available at
XXXXXXXX. Digital Object Identifier XXXXXX/TR.2023.XXXXXXX}}

\maketitle
\begin{abstract}
Adversarial attacks play a pivotal role in testing and improving the reliability of deep learning (DL) systems. Existing literature has demonstrated that subtle perturbations to the input can elicit erroneous outcomes, thereby substantially compromising the security of DL systems. This has emerged as a critical concern in the development of DL-based safety-critical systems like Autonomous Driving Systems (ADSs). The focus of existing adversarial attack methods on End-to-End (E2E) ADSs has predominantly centered on misbehaviors of steering angle, which overlooks speed-related controls or imperceptible perturbations. To address these challenges, we introduce UniAda—a multi-objective white-box attack technique with a core function that revolves around crafting an image-agnostic adversarial perturbation capable of simultaneously influencing both steering and speed controls. UniAda capitalizes on an intricately designed multi-objective optimization function with the Adaptive Weighting Scheme (AWS), enabling the concurrent optimization of diverse objectives. Validated with both simulated and real-world driving data, UniAda outperforms five benchmarks across two metrics, inducing steering and speed deviations from 3.54$^{\circ}$ to 29$^{\circ}$ and 11 km/h to 22 km/h on average. This systematic approach establishes UniAda as a proven technique for adversarial attacks on modern DL-based E2E ADSs.

\end{abstract}




\section{Introduction} \label{sec:intro}

\IEEEPARstart{A}{utonomous} driving has brought about a transformative shift in the driving experience, showcasing the immense potential to significantly reduce accidents attributed to human errors and alleviate traffic congestion challenges \cite{trafficjam}. 

\begin{figure}
    \centering
    \includegraphics[scale=0.26]{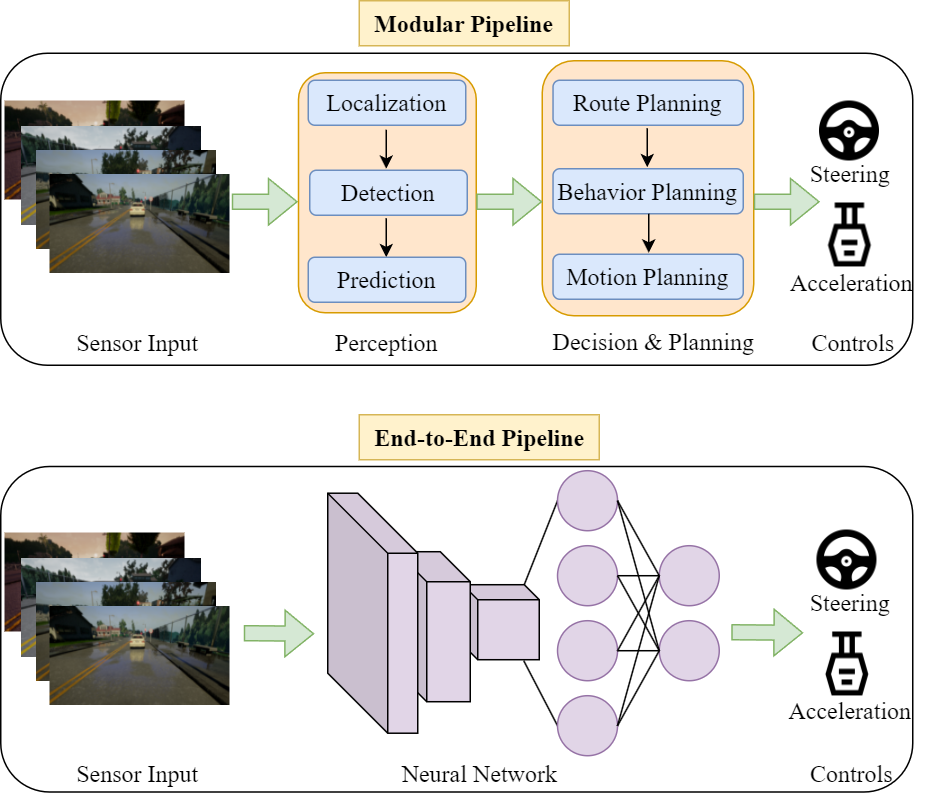}
    \caption{Overview of Modular ADS pipeline (top) and E2E Deep Learning-Based ADS pipeline (bottom)}
    \label{fig:modular&E2E ADS}
\end{figure}

There are two main approaches to implementing the ADS: modular \cite{yurtsever2020survey} and End-to-End (E2E) \cite{tampuu2020survey}. Fig. \ref{fig:modular&E2E ADS} gives a simple architecture overview of these two systems. The conventional modular approach decomposes the system into interconnected modules encompassing perception, decision-making, planning, and control \cite{modular1,modular2,yurtsever2020survey}. Recent advancements in deep learning have spurred interest in E2E methods, which combine multiple modules into a single deep neural network (usually CNN-based) model. E2E ADSs predominantly input sensor data (e.g., RGB images) and directly output numerical control actions such as steering and acceleration. While E2E ADSs have exhibited remarkable success \cite{dave, cil,exploring,udacity_dataset}, they remain susceptible to malicious input data, including misleading corner cases or adversarial examples \cite{deepbillboard, deeproad, physgan, deepxplore}.

Existing studies have delved into generating transformed or adversarial images as inputs to reveal the misleading prediction of the ADS under test \cite{deeproad,deepbillboard,deepxplore,physgan}. Usually, transformed images result from noticeable transformations applied to original inputs. Adversarial examples \cite{szegedy2013intriguing, fgsm,kurakin2016adversarial,BIM,tramer2017ensemble} involve subtle imperceptible perturbations to original data, avoiding human detection.
These studies explore diverse transformations or perturbations to the input images and analyze their impact on steering actions. These range from modifying weather conditions \cite{deeproad,li2021testing,deeptest} to introducing small black occlusions \cite{deepxplore}. However, these efforts fall short in comprehensively testing ADSs, marked by three limitations.

\begin{itemize}
  \item \textbf{Limited Vehicle Controls Testing}:
   Existing studies \cite{deepbillboard,von2023deepmaneuver,physgan,deeproad} mainly concentrate on attacking steering control, while neglecting speed-related controls. This falls short of testing the security of real-world driving. 
Notably, no literature exists on targeting image inputs solely to trigger errors in both steering and acceleration controls for E2E ADSs based on deep learning. 
  \item \textbf{Constrained Testing Scenarios}:
Current testing scenarios lack diversity. For instance, SelfOracle proposed by Stocco \textit{et~al.} \cite{selforacle} exclusively focuses on highway scenarios, excluding intersections, traffic lights, pedestrians, and other vehicles. Similarly, attack techniques like DeepBillboard \cite{deepbillboard} and PhysGAN \cite{physgan} concentrate on driving scenarios containing billboards. The state-of-the-art technique, DeepManeuver \cite{von2023deepmaneuver}, exclusively perform attacks within a simulator, lacking the testing of real-world scenarios. 
\item \textbf{Image-specific or Unnatural Perturbations}:
    Many existing techniques generate perturbations or transformations specific to individual images \cite{deeproad,deepxplore,deeptest,li2021testing}. Alternatively, they produce perturbations easily detectable by humans, such as unnatural billboard replacements \cite{deepbillboard,physgan,von2023deepmaneuver}. These limitations prompted our exploration of an adversarial perturbation—a subtle, human-imperceptible modification capable of consistently inducing errors in both speed and steering controls across a sequence of images (e.g., a driving record).
\end{itemize}

In addressing these shortcomings, we design \tool that leverages adaptive multi-objective learning to generate universal adversarial perturbations for ADSs. Notably, we term an adversarial perturbation as \textbf{universal} when it consistently triggers errors larger than a predefined threshold across most images sampled from the data distribution \cite{zhang2021survey,moosavi2017universal}, thereby transcending image-specific constraints. There could still exist a limited number of images within the distribution where the generated universal perturbation fails to trigger errors surpassing the predefined threshold. In our paper, we focus on the case where the distribution represents the set of images describing a driving scenario (e.g., car stops for a jaywalking pedestrian). To demonstrate the universality, we report the percentage of images that our generated perturbation attacks successfully by the Success Rate metric with four different tested thresholds for both objectives. The results have shown that UniAda can successfully attack 96.3\% images with steering error larger than $3.5^{\circ}$, which is a substantial improvement compared to baselines.

More specifically, UniAda optimizes a multi-objective function during perturbation generation. 
To achieve universality, \tool establishes a joint optimization function through the summation of multi-objective functions associated with each image within the input driving record.
Through iterative gradient descent, the joint function is minimized, facilitating the search for the perturbation. The resultant perturbation consistently triggers errors across multiple model predictions for most input images.
Departing from rudimentary uniform weighting or exhaustive grid search for objective weights, we propose the Adaptive Weighting Scheme (AWS), inspired by Chen \textit{et~al.} \cite{gradnorm}. AWS, a gradient-based weight adjustment method, strives to balance varying objective weights to ensure all objectives are trained at a similar rate. 
 This empowers UniAda to iteratively determine the optimal weight combination, enhancing the acquisition of shared feature representations for diverse objectives. To test the effectiveness of AWS, we proposes a variant of \tool, dubbed UniEqual, which shares the identical algorithm concepts but omits the use of AWS. We also conduct statistically analysis (i.e., $t$-test) to intuitively show the difference between \tool and UniEqual, which implies that AWS brings a statistically significant improvement in most cases.

We evaluate UniAda on 14 urban driving videos selected from four datasets (i.e,, Carla100 \cite{exploring}, Kitti \cite{geiger2013vision}, Udacity \cite{udacity_dataset}, Dave \cite{Dave_testing_dataset}) for two objectives (Steering and Acceleration) with three victim ADSs (i.e, CILRS, CILR, MotionTransformer). Our results demonstrate UniAda's superiority over five benchmarks (DeepManeuver \cite{von2023deepmaneuver}, DeepBillboard \cite{deepbillboard}, Perturbation Attack \cite{zhang2021evaluating}, FGSM \cite{fgsm}, UniEqual) in terms of key evaluation metrics—Mean Error and Success Rate. Notably, UniAda achieves the highest mean error in steering ($29.2^{\circ}$ for CILRS, $25.6^{\circ}$ for CILR, and $3.54^{\circ}$ for MotionTransformer), which is $6.6^{\circ}$, $8.6^{\circ}$ and $2.93^{\circ}$ higher than the single-objective state-of-the-art method DeepManeuver. Furthermore, compared to multi-objective techniques (i.e., Perturbation Attack, FGSM and UniEqual), UniAda emerges as the most effective solution for both objectives on average for all three models in almost all cases.



We summarize the key contributions of this paper as follows:

(1) We propose UniAda, a novel attack technique for DL-based E2E ADSs that can generate multi-objective universal adversarial perturbations for the input driving video. UniAda is effective in triggering errors on three state-of-the-art DL-based E2E ADSs for both steering and acceleration controls tested with both simulated and real-world data. To the best of our knowledge, we are the first to conduct imperceptible image-agnostic offline attacks to induce multi-objective misbehaviors for E2E ADSs.

(2) We propose a new strategy, Adaptive Weighting Scheme (AWS), employed in UniAda for balancing different objectives in a real-time manner, which maintains consistent training rates for each objective. The effectiveness of AWS is validated by comparing \tool with the equal-weighted counterpart, UniEqual.

(3) We validate the effectiveness of the multi-objective attack, by comparing it with single-objective counterparts. Additionally, we assess UniAda's effectiveness in urban traffic under both simulated and real-world driving environment. 

\section{Related Work}
\label{Background}
\subsection{Deep Learning in Autonomous Driving}
The integration of deep learning into E2E autonomous systems \cite{kuutti2020survey, chitta2022transfuser,casas2021mp3} traces back to the late 1980s, Pomerleau \textit{et~al.} \cite{pomerleau1988alvinn} built the Autonomous Land Vehicle in a Neural Network (ALVINN) system which uses a shallow three-layer network. Later in 2016,  Bojarski \textit{et~al.} \cite{dave} trained a Convolutional Neural Network (CNN) to directly map raw image pixels from a single front-facing camera to steering commands. Similarly, in the context of the Udacity self-driving car challenges \cite{udacity_dataset}, researchers embarked on constructing cutting-edge agents that interpret sensor inputs to govern vehicle operations. More recently, Codevilla \textit{et~al.} \cite{exploring} introduced an innovative conditional imitation learning framework, which takes RGB images, speed measurements and navigation instructions as inputs, generating steering and acceleration-related car controls as outputs.




\subsection{Attack Deep Learning Systems}

In the realm of contemporary software development, the widespread integration of deep learning has yielded satisfactory outcomes. However, extensive research \cite{deng2021deep, BIM, blackbox2, fgsm, physgan, liu2022attacking, lin2020adversarial, qi2022detection,woodlief2022semantic, pavlitskaya2020feasibility} has demonstrated that many deep learning models remain susceptible to intentional adversarial attacks designed to provoke misbehavior in these models. A prevalent form of such attacks involves the use of adversarial examples, characterized by imperceptible or quasi-imperceptible perturbations from the original inputs. By feeding the adversarial examples to the deep learning model, the model can produce different predictions from the original ones. This explores and reveals the vulnerabilities of the deep learning systems.

Out of all the vulnerabilities, the \textit{offline white-box attack} poses the most significant threat to deep learning systems, due to the full access to the targeted model and potential iterative interactions with it \cite{yuan2019adversarial}.

Both offline attacks and white-box attacks are categorized within the realm of adversarial attacks. Offline attacks \cite{codevilla2018offline, fgsm, haq2020comparing,zhang2021evaluating} apply the adversarial perturbation on a pre-acquired fixed dataset, where the attack performance is commonly evaluated by the prediction error between original and adversarial prediction. Specifically, offline attacks exploit the vulnerabilities without requiring a real-time interaction with the target system, allowing attackers to carefully analyze and manipulate models to design the attack technique. 




In the field of white-box attacks \cite{liu2009covering, nidhra2012black}, information regarding internal model structures and parameters are available for the attack techniques to trigger erroneous behavior, primarily beneficial for software verification. This in-depth understanding enables the attacker to formulate highly targeted and effective adversarial perturbations to manipulate the model's behavior. In the context of attacking DNNs, white-box attacks normally modify original inputs by using gradients computed with respect to relevant metrics, thereby inducing erroneous prediction (e.g., misclassification). For example, FGSM \cite{fgsm} utilizes gradient-based white-box attacks in image classification tasks such that the perturbed image will be wrongly classified, e.g., a perturbed dog image will be misclassified as cat by the DNN model with high accuracy.

In this paper, we focus on an offline white-box gradient-based attack strategy, which generates adversarial perturbations aimed at inducing misbehavior in DNN-based End-to-End autonomous driving systems.

\subsection{Offline Attacking Autonomous Driving Systems}
In the domain of offline attacks on DNN-based Autonomous driving systems, a plethora of methodologies has been proposed. Many of these approaches, such as \cite{deepbillboard,deeproad,deepxplore,physgan,li2021testing}, are designed to generate adversarial images as sensor inputs, with the objective of triggering errors in the targeted ADSs. These methods seek transformations from original images that result in transformed or adversarial images, capable of inducing model misbehavior. For instance, DeepXplore \cite{deepxplore} employs neuron coverage and differential behavior-based metrics within a white-box testing framework to create transformed images that mislead the steering angle predictions of ADSs. These transformed images might exhibit different lighting conditions or incorporate small black occlusions when compared to the originals. 


The aforementioned studies primarily focus on image-specific transformations, in which they apply distinct transformations to different images, rendering a single transformation ineffective across multiple images. In recent work, some studies \cite{deepbillboard, physgan, wu2023adversarial, von2023deepmaneuver} have pushed beyond this constraint by seeking transformations capable of consistently inducing model misbehavior, affecting sequences of images (i.e., driving records). Notably, Von \textit{et~al.} (DeepManeuver) \cite{von2023deepmaneuver}, Zhou \textit{et~al.} (DeepBillboard) \cite{deepbillboard} and Kong \textit{et~al.} (PhysGAN) \cite{physgan} utilize variants of prediction differences as target metrics to generate adversarial perturbations. These methods identify a universal perturbation pattern that can persistently provoke misbehavior on steering predictions across a sequence of driving images. However, their focus is confined to specific testing scenarios involving billboard appearances and targeting steering angle misbehavior exclusively. Moreover, DeepManeuver exclusively performs attacks within a simulator. 


The majority of these techniques concentrate on single-objective attacks on E2E ADSs, primarily aiming at misleading steering predictions. Beyond steering angle, speed-related controls bear equal significance in driving scenarios. This paper introduces a multi-objective attack strategy for E2E ADSs, generating perturbations that remain imperceptible to human eyes, yet consistently elicit model misbehavior across multiple controls. We also propose an adaptive weighting scheme to balance the influence of each objective, ensuring comparable training rates and higher efficiency in finding complementary perturbation patterns.


\begin{figure*}[tp]
    \centering
    \includegraphics[scale=0.165] {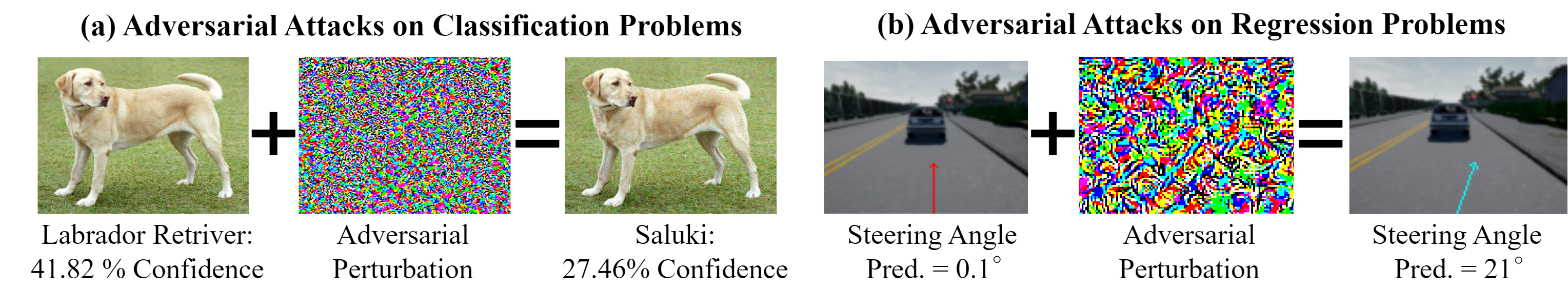}
    \caption{Examples of adversarial attacks on image classification and regression tasks. Pixel values of perturbation $\tau$ are scaled for visibility.}
    \label{fig:perturbation}
\end{figure*}

\section{Preliminaries} \label{sec: prob def}


\textbf{Adversarial Attacks:} In this section, we provide a concise and formulated overview of adversarial attacks, covering both image classification and regression problems. Additionally, we elucidate the concept of universal perturbation.

In adversarial attacks, the generated perturbation $\tau$ can trigger errors in the model prediction. By adding a subtle $\tau$ to the input sample $x$ (e.g., an RGB image), a model $m$ trained with optimal parameter $\mu$ can make a different prediction (mostly wrong) from the original. Specifically, for image classification problems, we wish to find a vector $\tau$ that can induce the CNN model to predict different class labels for $x$ and $x+\tau$, satisfying:

\begin{equation}
 m(x + \tau; \mu) \neq m(x; \mu)  \hspace{5mm}  s.t. \hspace{1mm} \Vert \tau \Vert_p \leq  \epsilon
\end{equation}
where $\Vert \cdot \Vert_p$ indicates the $L_p$ norm, and $L_0, L_2$ and $L_{\inf}$ are commonly used metrics \cite{moosavi2017universal, fgsm}. The constraint is to ensure $x+\tau$ and $x$ are visually the same to human eyes. For regression problems, the goal of the adversarial attack is to find $\tau$ that maximizes the discrepancy between the two numerical predictions, satisfying:

\begin{equation}
   \argmax_{\tau} \hspace{1mm} | m(x + \tau; \mu) - m(x; \mu) | \hspace{5mm} s.t. \hspace{1mm} \Vert \tau \Vert_p \leq  \epsilon
\end{equation}
Normally, in regression problems, the attack is considered successful if the prediction difference is larger by a predefined threshold \cite{deng2020analysis}.

Fig. \ref{fig:perturbation} gives examples of adversarial attacks on the image classification model (Fig. \ref{fig:perturbation}(a) with $\epsilon=15$) and autonomous driving regression model (Fig. \ref{fig:perturbation}(b) with $\epsilon=2$). Specifically, for the classification task, the CNN model is very vulnerable to a subtle perturbation, which changes its prediction from \textit{Labrador Retriver} with $41.82\%$ confidence to \textit{Saluki} with $27.46\%$ confidence. For the regression task, we show an example of a CNN-based E2E autonomous driving model that predicts the steering angle value from the image. The added imperceptible perturbation changes its prediction from $0.1^{\circ}$ to $21^{\circ}$. The above example shows that CNN models are vulnerable to such attacks.

For a \textbf{Universal Perturbation} \cite{moosavi2017universal} (i.e., image-agnostic), the goal is to seek a numerical vector $\tau$ that can successfully fool almost all datapoints sampled from $\mathbf{X}$, where $\mathbf{X}$ denotes a distribution of data. In regression problems, a universal perturbation $\tau$ satisfies:
\begin{equation}
\begin{aligned}
   &  |m(x + \tau; \mu) - m(x; \mu)| > \delta  \\
   & s.t. \hspace{1mm} \Vert \tau \Vert_p \leq  \epsilon \hspace{5mm} \textit{for most} \hspace{1mm} x \in \mathbf{X}
\end{aligned}
\end{equation}
In our paper, $\mathbf{X}$ is a distribution of images, represented by a driving record. We attack three CNN-based autonomous driving models (introduced in Section \ref{sec: ADS model}) that perform a multi-output regression task, which takes an RGB image as input and outputs continuous numerical steering angle and acceleration prediction.

\section{Threat Model}
Our work focuses on attacking DNN-based autonomous driving models in the image regression domain. We consider four types of threat model in this work: adversarial falsification, adversary's knowledge, adversarial specificity, and attack frequency \cite{yuan2019adversarial}.

\textbf{Adversarial Falsification:}
For intentional attack, the attacker can design an adversarial example $(x + \tau)$ to fool the model $m$ into making a wrong decision. However, this manipulated example is not visually different from $x$ to a human observer. For unintentional attack, the attacker can unintentionally send an input $\hat{x}$ that is sampled from a distribution different from the training distribution. Specifically, $\hat{x}$ has a  label $y$, and $m(\hat{x}) \neq y$. In this case, the model fails on $\hat{x}$ due to the distribution shift.

Our goal is to perform \textit{intentional} attack on regression problem, which adds a subtle perturbation $\tau$ on sample image $x$ to cause misbehavior on autonomous driving model $m$. The adversarial image is similar to the original image and conforms to the training distribution. 

\textbf{Adversary's Knowledge:} Based on the available knowledge to the model, attacks can be categorized into white-box, grey-box and black-box. They correspond to full access, partial access and no access to the model internal structure and parameter information. 
In our paper, we perform \textit{white-box} gradient-based attacks, where we know the model architecture, hyperparameters, model weights. We generate adversarial examples by calculating model gradients with our designed objective function. 

\textbf{Adversarial Specificity:} This includes targeted attacks and non-targeted attacks. For targeted attacks, the attacker misguides DNN to a specified prediction output (e.g., specified predicted class for classification problem, bigger/smaller predicted values for regression problems). In non-targeted attacks, the adversarial output can be arbitrary except the original one.

We construct a \textit{targeted} adversarial sample that follows specified attack direction. For example, we ask \tool to construct an adversarial sample that can induce the model to turn right (i.e., to predict bigger steering value) instead of turn left. 

\textbf{Attack Frequency:} This contains one-step attacks and iterative attacks, which differs from the number of interactions with the victim model. We use \textit{iterative} attacks that take multiple times to update the adversarial examples to reach a better performance.

\begin{figure*}[t]
    \centering
    \includegraphics[scale=0.25]{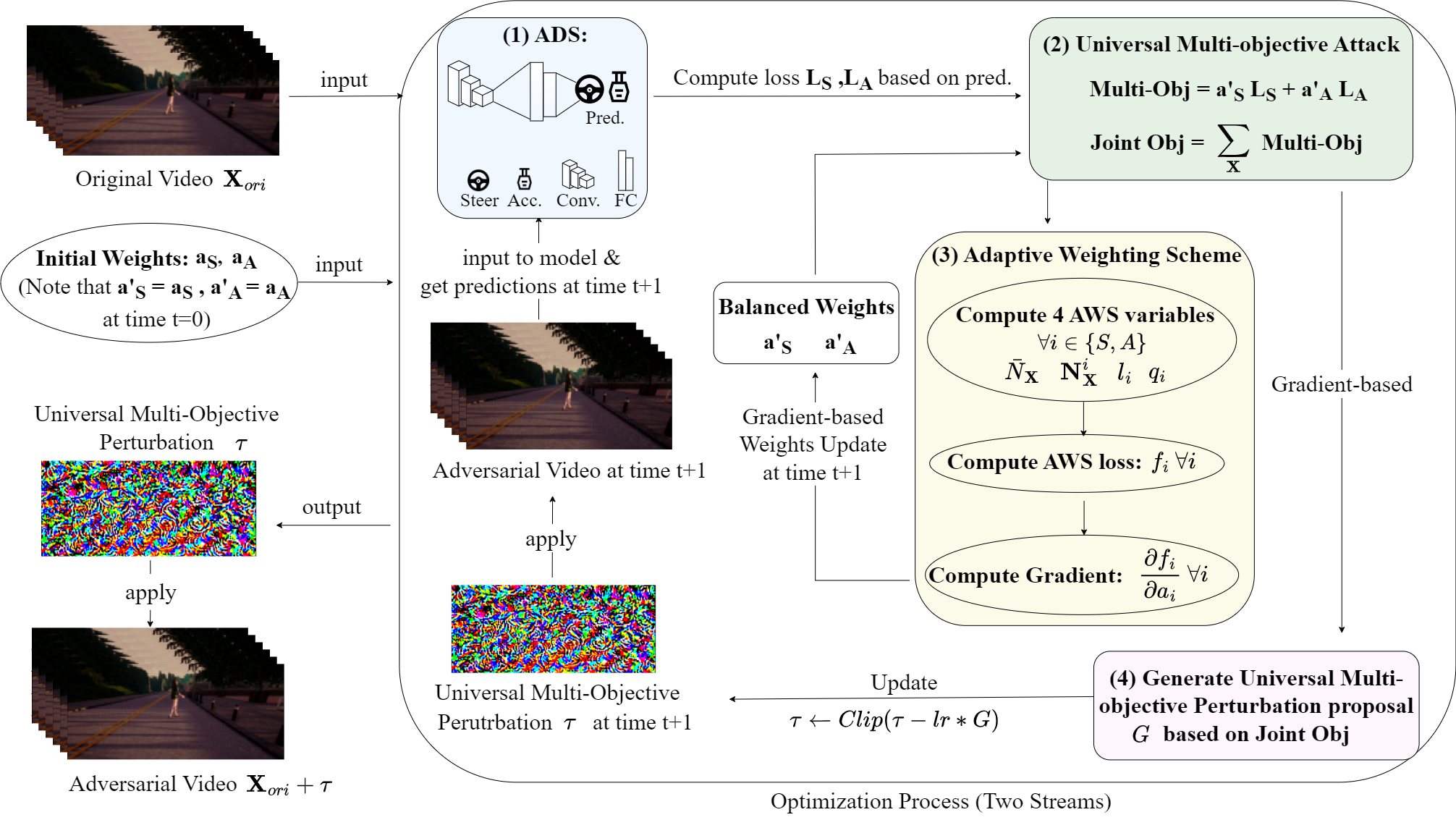}
    \caption{UniAda Overview: Our attack technique contains two main optimization streams. The first one is the Universal Multi-objective Attack (in green) to generate a perturbation that is influential to multiple images and multiple car controls (i.e., steering $S$ and acceleration $A$). The second one is the Adaptive Weighting Scheme (in yellow), which iteratively balances the objective weights to ensure similar training rates. Best viewed in colour.}
    \label{fig:workflow}
\end{figure*}
\section{Methodology}
\subsection{UniAda Overview}
This section outlines \tool, which employs adaptive objective weights to create a universal adversarial perturbation. This perturbation consistently misleads the targeted model across multiple images, impacting several car controls simultaneously.
The overview of \tool is shown in Fig. \ref{fig:workflow}. Given a driving video encompassing dozens of images and initial objective weights, \tool commences the perturbation discovery through parallel optimization streams: Perturbation Optimization and Objective Weight Optimization. These streams evolve in tandem during each iteration step $t$. 

In the perturbation optimization stream, each image inside the input video is fed into the autonomous driving model. This yields predictions for each car control, specifically steering angle and acceleration, as depicted in step \textit{(1) ADS}. Utilizing these predictions, \tool computes individual losses for each car control ($L_S, L_A$). Drawing from these losses and the prevailing objective weights $a'_S, a'_A$, \tool in step \textit{(2)} computes the joint optimization function (detailed explanations can be found in Section \ref{ssection:uni multi-obj test}). Subsequently, employing a gradient-based methodology, \tool derives the joint function's derivative with respect to the image, and the resultant gradient $G$ serves as the current perturbation proposal. $G$ is then leveraged to update the universal multi-objective perturbation for the subsequent time step $t+1$.

Simultaneously, the weight optimization procedure starts, enabling the dynamic adjustment of weights for each objective. At a given step $t$, \tool employs the ongoing objective weights $a'_S, a'_A$, coupled with the losses ($L_S, L_A$), to calculate four variables within step \textit{(3) Adaptive Weighting Scheme}. These variables contribute to the AWS loss. Subsequently, \tool computes gradients of the AWS loss $f_i$ with respect to each objective weight $a_i$. These gradients facilitate a gradient-based update mechanism, resulting in refreshed balanced weights at the following time step. Details for Adaptive Weighting Scheme are elaborated in Section \ref{ssection:aws}.

To generate universal multi-objective perturbations, both optimization processes iteratively proceed until the maximum number of epochs is reached. 

\subsection{Universal Multi-objective Attack} \label{ssection:uni multi-obj test}

We design the loss function \cite{physgan} for each objective $i$ to maximize the difference between original and adversarial model predictions:
\begin{equation}
\begin{aligned}
    L_i(\mathbf{X}^n_{ori},\tau) = & \frac{1}{\beta} \exp{- \frac{1}{\beta} (m_i(\mathbf{X}^n_{ori} + \tau) - m_i(\mathbf{X}^n_{ori})) \times d_i} \\
    & \forall n \in \{1,...,N\}
\end{aligned}
\label{eqn:singleloss}
\end{equation}
The hyperparameter $d_i = \pm 1$ is the attack direction for objective (car control) $i$, which is set to be $-1$/$1$ if we would like to attack in the negative/positive direction (adversarial prediction is learned to be smaller/larger than the original). 
Notations encompass $\tau$ for perturbation at time $t$, $\mathbf{X}^n_{ori}$ signifying the $n^{th}$ image frame inside the input video $\mathbf{X}_{ori}$, and $m_i(\mathbf{X}^n_{ori} + \tau)$ representing the prediction of autonomous driving model $m$ for objective $i$ from the perturbed adversarial image $\mathbf{X}^n_{ori} + \tau$. $\beta$ signifies the sharpness parameter. $N$ denotes the total number of images of the input driving record. Similar to DeepBillboard, we use original prediction as the attack reference.


Notably, prior work solely concentrated on perturbations inducing deviations in the model's steering angle predictions, disregarding speed-related car control objectives. To generate a universal perturbation capable of concurrently influencing multiple car controls, we devise a multi-objective loss as follows:
\begin{equation}
    O(\mathbf{X}^n_{ori}, \tau) = \sum_{i=1}^{C} a_i  L_i(\mathbf{X}^n_{ori},\tau) \hspace{3mm} \forall n \in \{1,...,N\}
\label{eqn:multi-objective loss}
\end{equation}
where $C$ denotes the number of objectives, and parameter $a_i$ reflects the weight assigned to objective $i$. In our case, there are two objectives, steering and acceleration, denoted by $i=S$ and $i=A$, respectively. Diverging from a fixed $a_i$, \tool introduces the Adaptive Weighting Scheme for real-time updates to $a_i$, effectively balancing each objective's contribution.

To generate a universally applicable perturbation that consistently impacts all input images, \tool find $\tau$ by optimizing the subsequent joint optimization function: 
\begin{equation}
    \argmin_{\tau} \sum_{n=1}^N  O(\mathbf{X}^n_{ori}, \tau)
\label{eqn:joint training loss}
\end{equation}

The generation of perturbation integrates the joint objective for all image frames within the video. Utilizing a gradient-based technique, \tool minimizes this joint function, iteratively refining $\tau$. 


\subsection{Adaptive Weighting Scheme} \label{ssection:aws}
During the multi-objective optimization process $O =\sum_i^C a_iL_i$ (as depicted in Equation \ref{eqn:multi-objective loss}), it is critical to select an appropriate weight for each objective. In our context, they significantly affect the contribution of each individual objective to the perturbation proposal. In the simplest case, one may opt for equal weighted objectives, where each objective is assigned a weight of $a_i = \frac{1}{C}$. In this case, the optimization process may not capture the importance of each objective accurately. Another option involves using computationally expensive grid search methods \cite{domingos2012useful}, which systematically explore a predefined set of values for each objective weight and select the combination that maximizes performance. However, it becomes computationally impractical for problems with a large parameter space. In our case, we employ an adaptive method \cite{gradnorm} dubbed Adaptive Weighting Scheme (AWS), where $a_i$ can vary at each time step $t$ during the perturbation optimization process. AWS is a complementary component integrated into the multi-objective optimization process, allowing us to determine the optimal value for each $a_i$ at each time step $t$. This ensures a balance in the contribution of each objective for optimal perturbation searching.
Specifically, AWS updates each objective weight based on its gradient to the AWS loss (\autoref{eqn:weightloss}) at each step $t$. AWS operates by assigning larger weights to objectives contributing less (indicating under-training) and smaller weights to those that are overly trained. Through AWS weight updates, both objectives collectively contribute to the perturbation, ensuring effective triggering of errors in both steering and acceleration controls, rather than confining influence to just one.

The AWS loss is formulated to capture differences between the weighted gradient norm for each objective $i$ (reflecting objective $i$'s perturbation pattern's impact on the universal perturbation) and the target norm (representing the desired influence of objective $i$'s perturbation on the universal perturbation). This target influence considers weighted gradient norms of other objectives ($\overline{N}_{\hat{X}}$) and the extent of misbehavior in objective $i$ compared to other objectives $(q_i)^{\gamma}$. The AWS loss $f_i$ for each $i$ is expressed as follows:
\begin{equation}
    f_i = \abs{N^i_{\hat{X}} - \overline{N}_{\hat{X}} \times (q_i)^{\gamma}}
\label{eqn:weightloss}
\end{equation}
where $f_i$ consists of four variables, $N^i_{\hat{X}}$, $\overline{N}_{\hat{X}}$, $l_i$, $q_i$, with $l_i$ utilized in the computation of $q_i$. These four variables represent the objective-specific norm, objective-average norm, the loss ratio that captures the inverse training rate for $i$, and the relative inverse training rate for $i$, respectively. The hyperparameter $\gamma$ controls the strength of rebalance; larger $\gamma$ exerts greater restoring force to pull objectives back to similar training rates.
Detailed formulas for each variable are presented below. 

\begin{algorithm}[t]
\footnotesize
\caption{Adaptive Weighting Scheme.}
\label{alg:AWS}
\begin{algorithmic}[1]
 \Require $\hat{X}$ - mini-batch images at search time $t$
 \Require $a_i$ - the weight of objective $i$ at time $t$
 \Require $L_i^0$ - initial loss for objective $i$
 \Require $lr_{grad}$ - hyperparameter, learning rate for AWS update
 \Require $\gamma$ - hyperparameter, strength of rebalance

         \For{$x$ \texttt{in} $\hat{X}$} 
             \State Compute $L_i(x), \frac{\partial L_i(x)}{\partial x} \hspace{2mm} \forall i$                   
             \EndFor
    
        \State Compute AWS variables $N^i_{\hat{X}}, \overline{N}_{\hat{X}}, l_i, q_i, f_i^{grad} \hspace{2mm} \forall i$ at time $t$
        
        \State Conditionally Update $a_i \gets a_i - lr_{grad} \times f_i^{grad}$
        \State Normalize $\sum_i a_i = 1$
  \end{algorithmic}
\end{algorithm}

1. $N^i_{\hat{X}} = E_{\hat{X}} \left[\Vert a_i \frac{\partial L_i(x)}{\partial x} \Vert_2 \right]$: the objective-specific norm\footnote{$\Vert \cdot \Vert_2$ denotes Euclidean norm, it is computed after flattening the three-dimensional input matrix to a 1D vector.}. $\hat{X} \subseteq \mathbf{X}_{ori} + \tau$ denotes the mini-batch image set at the time $t$, and $x \in \hat{X}$. $N^i_{\hat{X}}$ denotes the average of weighted gradient norms over all images inside the mini-batch at time $t$. We treat each image equally when computing the average. 


2. $\overline{N}_{\hat{X}} = E_i \left[ N^i_{\hat{X}}  \right]$: an objective-average norm that averages the objective-specific norms across all objectives.

3. $l_i = \frac{E_{\hat{X}} [L_i(x)] }{L_i^0}$: the loss ratio that captures the inverse training rate for the objective $i$. A smaller value indicates a higher training rate. Numerator is the batch average loss for objective $i$ at time $t$. The denominator denotes the initial loss for the objective $i$, which is the mean loss of the original prediction of $i$ at time zero over all images in the given video. The ratio is used to deal with different loss scales. With the inclusion of this variable, the objective weights are adjusted according to the misbehavior performance of $i$ for the current mini-batch images.

4. $q_i = \frac{l_i}{E_i[l_i]}$: the relative inverse training rate for the objective $i$, where $E_i[l_i]$ is the mean loss ratio over different objectives. A smaller $q_i$ gives UniAda a hint that the objective $i$ is trained too much (i.e., loss $L_i$ decreases too fast) compared to others, in which AWS will decrease its weight. 

Algorithm~\ref{alg:AWS} summarises the AWS process. 
For each image, AWS computes objective loss $L_i(x)$ and its gradient (Lines 1-3), retaining them for AWS variable computation (Line 4). Line 5 performs objective weights update if the new weight is positive. The positive weight constraint is grounded in the necessity for $L_i(x)$ and $O(x)$ to be positively correlated, ensuring the minimization of $L_i \hspace{2mm} \forall i$ during $O(x)$ minimization. Finally, the updated weights are normalized to ensure their summation equates to 1 (Line 6).

\subsection{Generating Adversarial Perturbations}


The above two sections constitute two main optimization streams: the perturbation optimization by the joint objective (\autoref{eqn:joint training loss}) in the universal multi-objective testing, and the objective weight optimization by AWS. Both streams operate in a gradient-based manner and are iteratively updated at each step $t$.
The algorithm details are summarized in Algorithm \ref{alg:uniada}, structured across four main stages: Input, Initialization, Perturbation Searching (comprising two optimization streams), and the Output stage. We elaborate on the details below.

\textbf{Input:} The algorithm requires four key inputs outlined in the \textbf{Require} lines. These inputs encompass the input video $\mathbf{X}_{ori}$, the autonomous driving model $m$ responsible for predicting values across multiple car controls, the attack direction $d_i$ corresponding to each objective $i$, and hyperparameters $hyper$. Table \ref{hyperparameters} explains the role and selected values of each hyperparameter.

\begin{table}[tbp]
\footnotesize
\caption{UniAda Hyperparameters: describe the meaning of each hyperparameters and their values used in the experiments.}
\begin{center}
\begin{tabular}{ccc}
\toprule
\textbf{Notation} & \textbf{Description} & \textbf{Value} \\
\midrule
$lr$ & Perturbation searching learning rate & 0.2\\
\midrule
$lr_{grad}$ & AWS learning rate & 0.005 \\
\midrule
$bs$ & Batch size & 5\\
\midrule
$\beta$ & Sharpness parameter in multi-objective loss & 2\\
\midrule
$Epochs$ & Number of Epochs & 250 \\
\midrule
$\gamma$ & AWS parameter controlling rebalance strength & 10\\
\midrule
$\theta$ & Perturbation gradient rescale threshold & 0.3\\
\midrule
$\epsilon$ & Maximum perturbation to the image & \{2, 5\}\\
\bottomrule
\end{tabular}
\end{center}
\label{hyperparameters}
\end{table}


\textbf{Initialization:} Lines 1-5 initialize all variables needed. Specifically, Line 1 creates a variable $\mathbf{X}$ to store all original image frames within the video. Line 2 initializes the universal multi-objective perturbation $\tau$ as a tensor of zeros, whose dimension is the same as the input image. Line 3 initializes objective weights $a_i$ uniformly. Line 4 computes initial loss $L^0_i$ for objective $i$, which is used to compute the AWS variable $l_i$. Lastly, Line 5 sets the training time $t$ as zero. 

\textbf{Perturbation Searching:} 
For each epoch, UniAda shuffles all video images (Line 7) and implements learning rate decay by a predefined schedule (reduced by a factor of 0.8 every 50 epochs, as standard in the DL community \cite{lrdecay}) to avoid local optima (Line 8). During each step $t$ of searching, UniAda iterates over mini-batches with batch size $bs$ (Lines 9-10). It subsequently computes single objective loss (Line 11), multi-objective loss (Line 12), and corresponding gradients with respect to each image (Line 13). Line 14 involves computation and storage of gradient norms used to determine AWS variable $N^i_{\hat{X}}$. After completing traversal across all mini-batch images in $\hat{X}$, UniAda generates gradient proposals $\frac{\partial O(x)}{\partial x}$ for each image $x$ in the mini-batch, alongside weighted gradient norms $\Vert a_i \frac{\partial L_i(x)}{\partial x} \Vert_2$ for each image and objective $i$. Line 16 calculates average gradient proposals across $\hat{X}$. In instances where gradients are too minimal to impact perturbation $\tau$, inducing sluggish search, UniAda rescales average gradients $G$ by $G = G \times \frac{\theta}{\Vert G \Vert_2} $ if $\Vert G \Vert_2 < \theta$ and $\Vert G \Vert_2 \neq 0$ (Line 17). This amplifies gradient magnitude, elevating its influence during addition to $\tau$. Subsequently, Line 18 updates total perturbation $\tau$ via gradient descent, enforcing a maximum perturbation size of $\epsilon$ to maintain imperceptibility. Afterward, all image frames within $\mathbf{X}$ undergo updates, subject to processing for perceptual fidelity (Line 19). This entails confining pixel values within the range [0, 255]. Lines 20-22 encapsulate the AWS procedure, as described in Algorithm \autoref{alg:AWS}. These steps iteratively unfold until the maximum epoch limit is attained.

\textbf{Output:} Upon reaching the maximum epoch count, UniAda yields a universal multi-objective perturbation $\tau$ for the input video. 

\begin{algorithm}[tbp]
\footnotesize
\caption{UniAda.}
\label{alg:uniada}
\begin{algorithmic}[1]
 \Require $\mathbf{X}_{ori}$ - input video
 \Require $m$ - the targeted autonomous driving model
 \Require $d_i \in \{+1, -1\}$ - Direction of attack: accelerate/decelerate, left/right
 \Require $hyper - \{lr, lr_{grad}, bs, \beta, Epochs, \gamma, \theta, \epsilon \}$, dictionary of input hyperparameters

  \State $\mathbf{X} = copy(\mathbf{X}_{ori})$   
  \State $\tau = zero(\mathbf{X}[0].shape)$           
  \State $a_i = \frac{1}{C} \hspace{2mm} \forall i \in \{1,...,C\}$
  \State Compute $L^0_i \hspace{2mm} \forall i$ 
  \State $t=0$
 
    \For{epoch in Epochs}
       \State random.shuffle($\mathbf{X}$)
       \State Adjust $lr$, $lr_{grad}$ for each 50 epochs 
       \For{$\hat{X}$ \texttt{in} $\mathbf{X}$}
         \For{$x$ \texttt{in} $\hat{X}$} 
             \State $L_i(x) = \frac{1}{\beta} \exp{-\frac{1}{\beta} (m_i(x)-m_i(x_{ori})) \times d_i} \hspace{2mm} \forall i$                         
             \State $O(x) = \sum_{i=1}^{C} a_i L_i(x)$     
             \State Compute $\frac{\partial O(x)}{\partial x}$
             \State Compute $\Vert a_i \frac{\partial L_i(x)}{\partial x} \Vert_2 \hspace{2mm} \forall i$
             \EndFor
        \State $G = \frac{1}{len(\hat{X})} \sum_{x \in \hat{X}} \frac{\partial O(x)}{\partial x}$                        
        \State Conditionally Rescale $G$ 
        \State Update $\tau \gets Clip_{\epsilon}(\tau - lr * G)$
        \State Update $\mathbf{X} \gets Process(\mathbf{X}_{ori} + \tau)$
        \State Compute $N^i_{\hat{X}}, \overline{N}_{\hat{X}}, l_i, q_i, f_i^{grad} \hspace{2mm} \forall i$
        
        \State Conditionally Update $a_i \gets a_i - lr_{grad} \times f_i^{grad}$
        \State Renormalize $\sum_i a_i = 1$
        \State $t = t+1$
        \EndFor 
      \EndFor
    \State \textbf{return} $\tau$  
  \end{algorithmic}
\end{algorithm}

\section{Experiments}
We evaluate UniAda on three widely-used multi-output autonomous driving models with 14 driving videos. The goal of our evaluation is to answer the following research questions.

\subsection{Research Questions}
\noindent\textbf{RQ1. UniAda Effectiveness:} \textit{How effective is UniAda in generating adversarial perturbations?}

To evaluate the effectiveness of UniAda in generating perturbations that trigger errors in multiple controls, we compare its performance on 14 driving videos (7 simulated and 7 real-world videos) with four existing methods (DeepManeuver, Perturbation Attack, DeepBillboard, FGSM) on two evaluation metrics (mean error and success rate).



\noindent\textbf{RQ2. Adaptive Weighting Scheme Effectiveness:} \textit{How can the Adaptive Weighting Scheme assist the attack performance?}

To assess the effectiveness of AWS, we introduce a variant of \tool dubbed UniEqual, wherein objectives are assigned equal weights during the perturbation search process. By comparing UniEqual with \tool, we can observe the exact impact of AWS. Furthermore, we graphically depict the trajectory of each objective weight to provide insights into the operation of AWS. We also conduct the statistical analysis ($t$-test) to show the significance of the improvement. 

\vspace{.2em}
\noindent\textbf{RQ3. Multi-objective Attack Effectiveness:} \textit{Does the multi-objective adversarial attack improve the attack effectiveness for all objectives simultaneously compared to the single-objective attack?}

Attacking multiple objectives together may yield a better result for each objective than single-objective attacks, since attacking diverse objectives can offer shared insights sometimes, yielding a more comprehensive and effective perturbation pattern.
To answer this question, we compare the effectiveness of the adversarial perturbation generated by the multi-objective loss (\autoref{eqn:multi-objective loss}) with those by single objective loss (\autoref{eqn:singleloss}).

\subsection{Experimental Setup}
This section provides a comprehensive overview of our experimental setup, encompassing datasets, victim autonomous driving models, baseline methods, evaluation metrics, and hyperparameters.

\subsubsection{Dataset}

Our experimentation is grounded in the use of 14 driving videos in total, with 7 simulated videos and 7 real-world videos. 
For the simulated videos, we extracted all 7 videos from the Carla100 dataset \cite{exploring}, which contains realistic simulated driving data (e.g., RGB images, steering and speed-related control data) captured by the central camera with 100 hours of driving from an urban town of the Carla simulator (version 0.8.4). For the real-world videos, we extracted: (1) Two videos from Dave testing dataset \cite{Dave_testing_dataset} which contains 45,568 real-world driving images to test the NVIDIA Dave model; (2) Three videos from Udacity self-driving car challenge dataset \cite{udacity_dataset}  that contains 101,396 real-world driving images captured by a dashboard-mounted camera of a human-driven car; (3) Two videos from Kitti dataset \cite{geiger2013vision} which contains 14,999 real-world driving images from six different scenes captured from a VW Passat station wagon equipped with four cameras. 


Our manual selection of videos captures various scenarios under urban traffic, such as different weather conditions and different driving maneuvers. 
Further details regarding these 14 videos are summarised in Table \ref{scenes}.

\begin{table}
\scriptsize
 \begin{adjustwidth}{-1.5mm}{}
\caption{Descriptions for the selected videos}
\begin{center}
\begin{tabular}{cc}
\toprule
\textbf{Videos (No. of Imgs)} & \textbf{Description}\\
\midrule
 Carla Pedestrian (18) & Stop for a pedestrian, clear sunset \\
\midrule
Carla White Car (36) & Stop for the front white car, wet noon \\
\midrule
 Carla Black Car (22) & Stop for the front black car, clear noon\\
\midrule
Carla Gray Car (42) & Stop for the front gray car, clear sunset \\
\midrule
 Carla Blue Car (43) & Driving Straight, following a blue car, clear sunset\\
\midrule
Carla Red Light (66) & Stop for the front red light, wet noon \\
\midrule
Carla Light-blue Car (21) & Stop for the front light-blue car, clear noon\\ 
\midrule
 Dave Curve1 (34) & Turning left at crossroad, clear noon \\
\midrule
 Dave Straight1 (54) & Driving straight, urban two-way road, clear noon \\
\midrule
 Udacity Straight1 (21) & Stop for a white SUV, urban one-way road, clear noon  \\
\midrule
 Udacity Straight2 (22) & Driving Straight, urban one-way road, clear noon \\
\midrule
 Udacity Curve1 (15) & Turning left, urban two-way road, dusk \\
\midrule
 Kitti Curve1 (21) & Turning Right, urban one-way road, clear noon \\
\midrule
 Kitti Straight1 (21) & Driving Straight, urban one-way road,  clear noon \\
\bottomrule
\end{tabular}
\end{center}
\label{scenes}
\end{adjustwidth}
\end{table}


\subsubsection{Autonomous Driving Model} \label{sec: ADS model}
In the assessment, we select three end-to-end autonomous driving models for testing: CILRS, CILR, and MotionTransformer. Table \ref{tab:performance table} lists the dataset used for training and validation and the performance of models. Following common practice \cite{deepxplore,kim2019guiding} in autonomous driving research, we report Mean Square Error (MSE) between actual and predicted values as a measure of model accuracy. For consistency, we compute MSE on steering in radians and acceleration in [0,1] for all models on their provided validation sets. 

CILRS and CILR are conditional imitation learning models, proposed by Codevilla \textit{et~al.} \cite{exploring}. CILR uses the ResNet perception module as the backbone to extract information from RGB inputs. CILRS extends CILR architecture with an additional speed prediction head to incorporate speed-related features into the representation. They achieve outstanding performance in Carla's simulated urban traffic scenarios. Both models are trained using 10 hours of expert demonstrations and validated on a 2-hour subset (extracted from the Carla100 dataset). For both models, we use public-available pre-trained weights \footnote{https://github.com/felipecode/coiltraine}. Readers can refer to the original work \cite{exploring} for training details (e.g., training objectives, hyperparameter settings). On the provided validation set, both models achieve 0.0002 MSE in steering and approximately 0.03 MSE in acceleration.

MotionTransformer (MT) \cite{oinar2022self} utilizes RGB and optical flow images to learn both position and motion information. It employs two ResNet backbones for feature extraction from RGB and optical flow images, followed by a transformer encoder to distill knowledge from both. We use public-available pre-trained weights provided by the original work \footnote{https://github.com/chingisooinar/AI\_self-driving-car}. During their training process (please refer to the original work \cite{oinar2022self} for training details), the first 90\% of samples from the Udacity self-driving car challenge dataset are used for training, and the remaining 10\% is used for validation. On the validation set, MT achieves 0.002 MSE in steering and 0.0079 in acceleration, surpassing the widely-used steering angle prediction model, NVIDIA's DAVE2 \cite{dave}, which only achieves 0.0449 MSE in steering based on our experiments. 


\begin{table}[]
    \centering
    \begin{tabular}{c  c  c c}
    \toprule
    \multirow{2}{*}{Train \& Val set} & \multirow{2}{*}{Model} & \multicolumn{2}{c}{Performance (MSE)} \\
       &  & Steering & Acceleration \\
        \midrule
        Carla100 & CILR & 0.0002 & 0.0382 \\
        \midrule
       Carla100 &  CILRS & 0.0002 & 0.0343 \\
        \midrule
         Udacity & MT & 0.0020 & 0.0079 \\
        \bottomrule
    \end{tabular}
    \caption{Model Performance (MSE) on the provided validation set. }
    \label{tab:performance table}
\end{table}

For the experimentation, only RGB input undergoes perturbation, while other inputs (if there exists) remain unchanged. To maintain consistency and simplicity, we convert the output steering values from radians into degrees ($^\circ$) and acceleration values into speed values (km/h) according to the data documentation.

\subsubsection{Baseline} \label{sec: baseline}
We test five baselines to evaluate the effectiveness of UniAda. To ensure a fair comparison, for all methods, we set a uniform maximum perturbation, we set the attack surface to the whole image, and we perform targeted adversarial attacks. We use the best hyperparameter values reported in their paper for experiments.

\textbf{Fast Gradient Sign Method (FGSM):} 
Proposed by Goodfellow \textit{et~al.} \cite{fgsm}, FGSM is a one-step gradient-based attack approach to craft image-specific adversarial perturbations. To facilitate a meaningful comparison with UniAda, we extend FGSM to perform targeted attacks on multiple car controls by employing an equal-weighted multi-objective loss (same loss as UniEqual) as the optimization criterion during the perturbation search.



\textbf{DeepBillboard (DB):} Introduced by Zhou \textit{et~al.} \cite{deepbillboard}, DeepBillboard represents an iterative state-of-the-art attack method. 
It performs targeted attacks (intentionally misleads the model to steer left or right) and consistently misleads the model's steering predictions. The adversarial perturbation pattern generation process is intrinsically linked to the constraint applied to the steering angle. Due to this inherent constraint, unlike FGSM, it is challenging to extend DeepBillboard to multiple objectives. Consequently, we present the steering angle results exclusively.


\textbf{DeepManeuver (DM)} \cite{von2023deepmaneuver} is a state-of-the-art gradient-based adversarial test generation approach for autonomous vehicles. It generates a perturbation patch on the predefined attack surface, consistently misleading the model’s steering prediction into the targeted direction (i.e., turn left or right). Similarly, we exclusively report steering results due to the design of its algorithm.

\textbf{Perturbation Attack (PA)} \cite{zhang2021evaluating} assesses the impact of image-specific perturbation attacks and patch attacks on the 3D object detector. The former applies the attack to the entire image while the latter achieves the attack by applying a small patch to the input image. To enable a meaningful comparison with UniAda, we utilize the perturbation attack and an equal-weighted multi-objective loss (the same loss as UniEqual, enabling the specifications of attack directions) as the optimization criterion to induce misbehavior in car controls instead of 3D bounding boxes.



\textbf{UniEqual:} A variant of UniAda, sharing an identical universal multi-objective attack algorithm. However, UniEqual assigns equal weights to different objectives, omitting the use of AWS. Similar to UniAda, it is an iterative attack method designed to generate a universal multi-objective perturbation.

\subsubsection{Evaluation Metrics}
In this section, we introduce the evaluation metrics utilized in our experimentation. To assess the attack effectiveness of the generated examples, we adhere to established practices \cite{deng2020analysis, deepbillboard} and employ two widely-used metrics for the offline evaluation of ADS models: Mean Error (ME) \cite{wu2023adversarial} and Success Rate (SR) \cite{deng2020analysis}, as defined below for each objective $i$:
\begin{equation}\label{eqn:ME}
    ME_i =\frac{1}{N} \sum_{n=1}^{N} [d_i \times (m_i(\mathbf{X}^n_{ori} + \tau) - m_i(\mathbf{X}^n_{ori}))]
\end{equation}

\begin{equation}\label{eqn:sr}
    SR_i = \frac{1}{N} \sum_{n=1}^{N} I(d_i \times (m_i(\mathbf{X}^n_{ori} + \tau) - m_i(\mathbf{X}_{ori}^n)) > \delta_i)
\end{equation}
The variables mentioned above are detailed in Section \ref{sec: prob def}. $I$ is an indicator function that returns 0 or 1. Notably, the evaluation metrics are designed to capture the power of the attack. Larger metric values signify more effective attacks. In order to punish incorrect attack directions (i.e., situations when the attacker wishes the car to turn right but the attack algorithm misleads the car into turning left), the use of sign-invariant operators such as absolute or squared values in the metrics is avoided.
 
Mean error $ME$ indicates the average strength of our attack on the input video. Success rate $SR$ provides a hierarchy of performance by segmenting results based on different thresholds. More specifically, the success rate measures the proportion of image frames in the video that have been successfully attacked. An attack is deemed successful if the discrepancy between the prediction made on the generated adversarial image and the original prediction surpasses a predetermined threshold. Thresholds for steering and acceleration are denoted as $\delta_{S}$ and $\delta_{A}$, respectively. By testing various threshold values, we gain hierarchical insights into how the universal adversarial perturbation impacts a sequence of model predictions.


\begin{figure}
    \centering
    \includegraphics[scale=0.186]{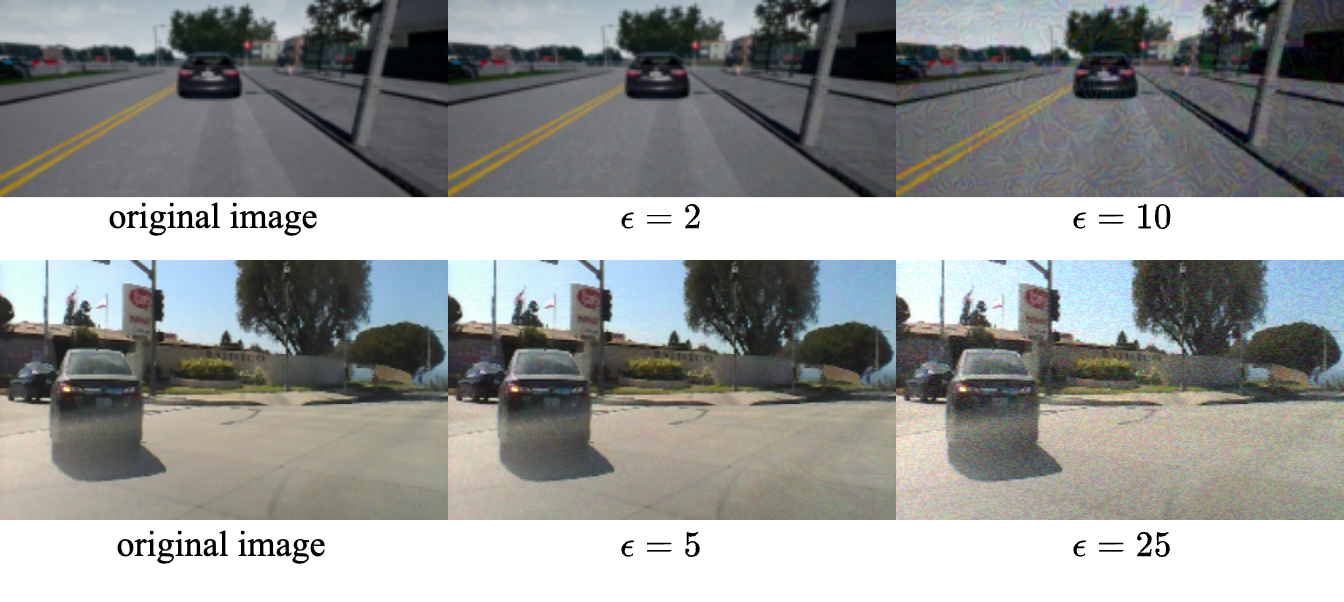}
    \caption{Adversarial images with different $\epsilon$ values.}
    \label{fig:eps visual}
\end{figure}

\begin{figure*}
    \centering
    \includegraphics[scale=0.3]{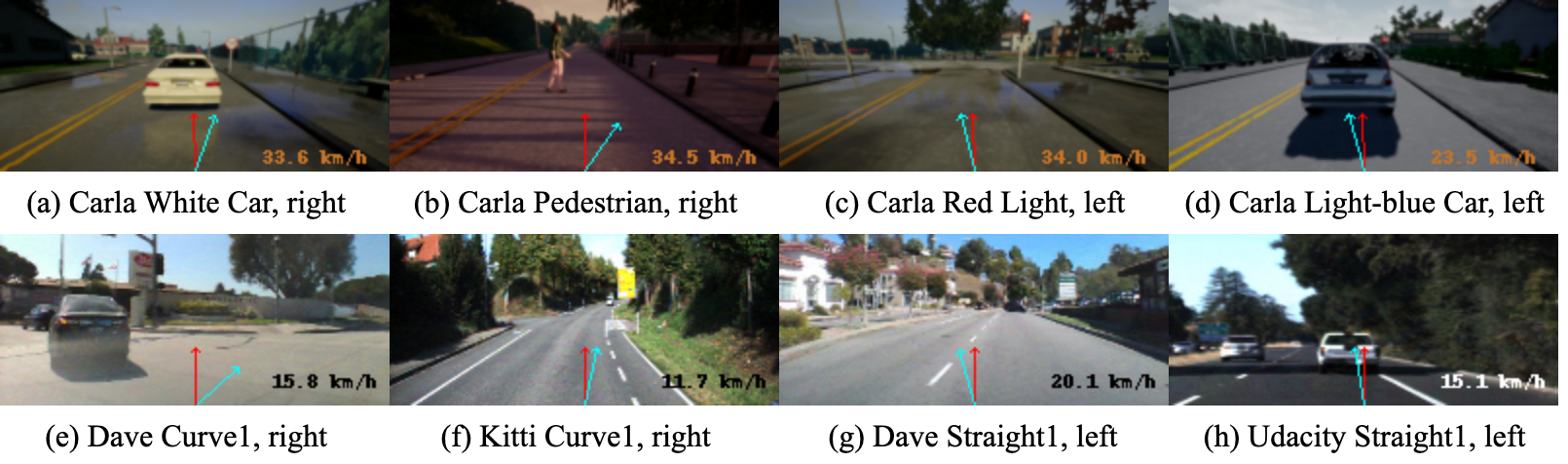}
    \caption{Adversarial Images generated by UniAda: The first/second row is adversarial images generated from simulated/real-world videos. Blue/Red arrows indicate the adversarial/original steering angle predicted from the image. The bottom right shows the acceleration error.}
    \label{sampleerrors}
\end{figure*}

\subsubsection{Hyperparameters}
For the baseline methods, we use the best hyperparameters reported in their paper for the experimentation. Hyperparameters for \tool are outlined and explained in Table \ref{hyperparameters}.

To ensure the generated perturbations remain imperceptible to humans, we set the maximum perturbation to $\epsilon$=2 for simulated videos and $\epsilon$=5 for real-world videos. This value is chosen to strike a balance between inducing misbehavior and maintaining the visual similarity between the original and perturbed images. In Fig. \ref{fig:eps visual}, we illustrate examples of adversarial images produced with perturbations of different $\epsilon$ values. Notably, for simulated images with $\epsilon$=2 and real-world images with $\epsilon$=5, the perturbed image closely resembles the original. For higher $\epsilon$ values, the distinction between an adversarial image and its original counterpart becomes more evident. 

\textbf{Reproducibility.} The trained model, framework, and data are available at https://github.com/UniAdaRepo/UniAda/.

\subsection{Results and Analyses}
In this section, we present our results and analyses to address the research questions. For all methods, the targeted attack directions are set to $d_S$=1 (steering to the right) and $d_A$=1 (acceleration). Positive/Negative error values indicate alignment/misalignment with the specified attack directions. All experiments are repeated five times and average results are reported. 

\vspace{0.3em}
\noindent\textbf{Visualization of Generated Adversarial Images:}
We provide visualizations of example adversarial images generated by UniAda in Fig. \ref{sampleerrors} for both simulated and real-world data. The top/bottom row displays adversarial images generated from simulated/real-world dataset. Red/blue arrows represent the original/adversarial steering angle prediction. The bottom-right corner of each image indicates the acceleration error. 

\begin{table*}
\robustify\bfseries
\scriptsize
\caption{Mean Error Result Table: mean error results for 7 simulated videos of all six techniques, tested with CILRS and CILR models. Last line indicates the average result over 7 videos. Best results are highlighted in bold. We approximate the floating steering angle to degrees $^{\circ}$ in $ME_S$ and floating acceleration to km/h in $ME_A$ for a more straightforward attack effect. - denotes unavailable results.}
\begin{center}
\begin{tabular}{cc| S[table-format=3.2, detect-all] ccccc | S[table-format=3.2, detect-all]  c S[table-format=3.2, detect-all, table-number-alignment=left]  ccc}
\hline
\multirow{2}{*}{Videos} & \multirow{2}{*}{Metrics} & \multicolumn{6}{c}{CILRS} & \multicolumn{6}{|c}{CILR}\\

&  & {UniAda} & DM  & DB & PA & FGSM & UniEqual & {UniAda}   & DM & {DB} & PA & FGSM & UniEqual\\
\hline
Carla & $ME_S$ & 39.9 & \bfseries 40.5 & 32.0 & 32.6 & 1.40 & 39.5 & \bfseries 30.1 & 23.9 & {12.7} & 22.5 & 0.20 & 30.0 \\
Pedestrian &  $ME_A$ & 22.9& $-$ & $-$ & 16.1 & 15.6 & \bfseries 23.2 & \bfseries13.3* & $-$ & {$-$} & 11.6 & 9.70 & 13.2 \\
\hline
Carla & $ME_S$ & \bfseries25.6* & 18.1 & 20.5 & 16.1 & 1.40 & 24.6 & \bfseries 19.6 & 10.4 & {9.79} & 11.1 &  0.90& 19.5 \\
White Car & $ME_A$ & \bfseries 22.0* & $-$ & $-$ & 20.7 &15.6  & 21.1 & \bfseries15.4* & $-$ & {$-$} & 14.6 &12.9 & 14.8 \\
\hline
Carla & $ME_S$ & \bfseries 18.6* & 6.25 & 5.47 & 13.5 & 0.70 & 9.63 & \bfseries 19.6* & 12.3 & {6.81} & 9.51 &1.20  & 13.8 \\
Black Car & $ME_A$ & 23.8  & $-$ & $-$ & 16.8 &8.70  & \textbf{23.9} & \bfseries 25.4* &$-$ & {$-$} & 18.6 &6.00  & 21.3 \\
\hline
Carla & $ME_S$ & \bfseries 23.7* & 11.8 & 5.88 & 12.0 & 0.70 & 7.08 & \bfseries 14.8* & 2.21 & {12.0} & 11.8 & 1.20 & 13.3 \\
Gray Car & $ME_A$ & \bfseries 17.4* & $-$ & $-$& 13.9 & 6.90  & 14.7 & \textbf{18.2} & $-$ & {$-$} & 16.5 & 5.06  & 18.1 \\
\hline
Carla & $ME_S$ & \bfseries 28.6* & 24.8 & 26.1 & 14.5 & 1.40 & 22.3 & \bfseries 31.2* & 18.5 & {9.68} & 15.1 & 0.10  & 30.7  \\
Blue Car & $ME_A$ & \bfseries 22.6& $-$ & $-$ & 21.9 & 8.74  & 22.5 & \bfseries24.3* & $-$ & {$-$} & 21.2 & 8.28 & 23.8  \\
\hline
Carla & $ME_S$ & \bfseries 41.9* & 38.9 & 35.4 & 21.2 & 0.70 & 41.2 & \bfseries 43.4* & 31.9 & 36.5 & 17.9 & 0.60 & 42.7 \\
Red Light & $ME_A$ & \textbf{26.3} & $-$ & $-$& 23.1 &11.0  & 26.2 & 26.3 & $-$& {$-$} & 21.8 & 9.20 & \textbf{26.4}  \\
\hline
Carla & $ME_S$ & \bfseries 25.9* & 17.5 & 15.0 & 6.37 & 0.70 & 11.2 & 20.4 & 19.6 & \bfseries 23.8* & 5.25 & 0.80 & 7.85\\ 
Light-blue Car & $ME_A$ & 21.0 & $-$ & $-$ & 19.9 & 8.74 & \bfseries 22.0 & \bfseries 14.0* & \text{--} & {$-$} & 0.91 & 0.92 & 4.16 \\
\hline
\multirow{2}{*}{Average} & $ME_S$ & \bfseries 29.2* & 22.6 & 20.1 & 16.6 & 1.00 & 22.2 & \bfseries25.6* & 17.0 & 15.9 & 13.3 & 0.71 & 22.6 \\
&$ME_A$ & \bfseries 22.3 & $-$ & $-$ & 18.9 & 10.8 & 21.9 & \bfseries19.6* & $-$ & {$-$} & 15.0 &  7.44 & 17.4 \\
\hline
\end{tabular}
\end{center}
\label{tbl:me result}
\end{table*}

\begin{table}[]
\robustify\bfseries
\scriptsize
 \begin{adjustwidth}{-1.5mm}{}
\caption{Mean Error Result for 7 real-world driving videos of all six techniques with MotionTransformer model under test. Avg. denotes the average results over 7 videos.}
    \centering
    \begin{tabular}{c| c| c ccccc}
    \hline
    \multirow{2}{*}{Videos} & \multirow{2}{*}{Metrics} & \multicolumn{6}{c}{MotionTransformer}  \\
    && {UniAda} & DM  & DB & PA & FGSM & UniEqual \\
    \hline
       Dave  & $ME_S$ & \multicolumn{1}{S[table-format=1.3, detect-all]}{\bfseries 7.25*} & 1.54 & 1.88 & 5.69 & 0.94 & 5.73 \\
        Curve1 & $ME_A$  & \multicolumn{1}{S[table-format=2.2, detect-all]}{\bfseries 14.3*} &$-$&$-$&5.21 & 2.54 & 3.97 \\
         \hline
        Dave & $ME_S$ & \multicolumn{1}{S[table-format=1.3, detect-all]}{\bfseries  1.01*} & -0.01 & 0.25 & 0.30 & -0.08 & 0.63 \\
        Straight1 & $ME_A$ & \multicolumn{1}{S[table-format=1.3, detect-all]}{\bfseries 4.68*} & $-$ & $-$ & -5.35 & -7.55 & -2.33 \\
        \hline
        Udacity & $ME_S$ & \multicolumn{1}{S[table-format=1.3, detect-all]}{ \bfseries  8.15} & 7.47 & 7.31 & 8.11 & 4.51 & 7.99 \\
        Straight1 & $ME_A$ & \multicolumn{1}{S[table-format=2.2, detect-all]}{\bfseries 16.5*}& $-$ & $-$ & 10.1 & 3.55 & 6.18 \\
        \hline
        Udacity & $ME_S$ & \multicolumn{1}{S[table-format=1.3, detect-all]}{\bfseries  2.75*} & -0.25 & -0.01 & 0.73 & 0.14 & 1.57  \\
        Straight2 & $ME_A$ & \multicolumn{1}{S[table-format=2.2, detect-all]}{\bfseries  10.3* }& $-$ & $-$ & -4.65 & -2.06 & 1.29 \\
        \hline
        Udacity & $ME_S$ & \multicolumn{1}{S[table-format=1.3, detect-all]}{\bfseries 1.97} & -3.09 & -2.18 & 1.07 & 0.83 & 1.94 \\
        Curve1 & $ME_A$ & \multicolumn{1}{S[table-format=2.2, detect-all]}{\bfseries  12.2*} & $-$ & $-$ & 3.00 &1.99 & -2.35 \\
        \hline
        Kitti & $ME_S$ & \multicolumn{1}{S[table-format=1.3, detect-all]}{ 2.90} & 2.76 & \textbf{3.14} & 2.90 & 0.84 & 3.10 \\
        Curve1 & $ME_A$ & \multicolumn{1}{S[table-format=1.3, detect-all]}{\bfseries  7.20*} & $-$ & $-$ & 1.41 & 0.34 & -0.59 \\
        \hline
        Kitti & $ME_S$ & \multicolumn{1}{S[table-format=1.3, detect-all]}{\bfseries 0.74*} & -4.16 & -2.45 & 0.09 & 0.54 & 0.63 \\
        Straight1 & $ME_A$ & \multicolumn{1}{S[table-format=2.2, detect-all]}{\bfseries  11.7*} & $-$ & $-$ & 1.01 & -1.65 & 1.75 \\
        \hline
        \multirow{2}{*}{Avg.} & $ME_S$ & \multicolumn{1}{S[table-format=1.3, detect-all]}{\bfseries  3.54* }& 0.61 & 1.13 & 2.69 & 1.10 & 3.08 \\
        & $ME_A$ & \multicolumn{1}{S[table-format=2.2, detect-all]}{\bfseries  11.0*} & $-$ & $-$ & 1.53 & -0.41 & 1.13 \\
        \hline
    \end{tabular}
    \label{tab:me real-world result}
\end{adjustwidth}
\end{table}
\vspace{0.3em}
\noindent\textbf{RQ1. UniAda Effectiveness.}
We evaluate the effectiveness of UniAda in comparison to four baselines: DeepManeuver, Perturbation Attack, DeepBillboard and FGSM, using two evaluation metrics across 14 videos and 3 ADSs.

\textbf{Mean Error Results:} 
Results for simulated and real-world data are presented in Table \ref{tbl:me result} and Table \ref{tab:me real-world result}. UniAda outperforms all four baselines in both steering angle and acceleration objectives in the average results for both simulated and real-world data. 
For simulated videos, under CILRS, UniAda causes an average steering error of approximately 29.2$^{\circ}$, which is 6.6$^{\circ}$, 9.1$^{\circ}$, 12.6$^{\circ}$ and 28.2$^{\circ}$ greater than DM, DB, PA, and FGSM, respectively. For acceleration, UniAda leads to an acceleration error of around 22.3 km/h, surpassing PA and FGSM by 3.4 km/h and 11.5 km/h, respectively. For CILR, \tool maintains its leading position for both objectives, causing an average steering error of 25.6$^{\circ}$, with a maximum error of 43.4$^{\circ}$ under ``Red Light''. For acceleration, it reaches an average error of 19.6 km/h. Among the four baselines, for both models on average, DM demonstrates the best performance, while FGSM performs the worst. DB and PA secure middle positions. Delving into the performance of individual videos, we found that under the CILR ``Light-blue Car'' video, UniAda is slightly outperformed by DB in $ME_S$ by around 3.4$^{\circ}$. Similarly, in CILRS ``Pedestrian'', \tool is slightly outperformed by DM by 0.6$^{\circ}$. However, DB and DM can only target one objective at a time, making them less functional compared to \tool. Furthermore, when considering other videos and the average result, \tool consistently outperforms them by a significant margin.



For real-world videos, \tool reaches the best performance on average, with 3.54$^{\circ}$ in $ME_S$ and 11.0 km/h in $ME_A$, which is about triple $ME_S$ of DB. Only in one case ("Kitti Curve1") DB slightly outperforms \tool by 0.24$^{\circ}$. On average, PA performs the best among four baselines, but is 9.47 km/h and 0.85$^{\circ}$ smaller than \tool. DM yields the worst performance, resulting in 0.61$^{\circ}$ in $ME_S$, which may be attributed to the fact that DM is designed for online testing within simulators.



\textbf{Targeted Attack Effectiveness:} From Table \ref{tbl:me result}, all methods demonstrate positive error values, indicating their effectiveness in misleading CILRS and CILR in the intended direction. However, when attacking the MT model (Table \ref{tab:me real-world result}), all methods except \tool occasionally generate adversarial samples that mislead the model into unintended directions, thereby failing to accomplish targeted attacks. Additionally, all techniques achieve a relatively low error compared to attacking CILRS and CILR. The weak performance in attacking the MT model may be attributed to the fact that, according to our settings, only the RGB input undergoes perturbations, whereas the MT model extracts features from both RGB and optical flow images. Out of 7 videos, FGSM induces MT to decelerate in 3 cases, and 2 cases for PA (i.e., negative $ME_A$, which contradicts the intended acceleration target). In teams of steering angle, FGSM incorrectly induces the MT to turn left instead of right as instructed in one case, while PA, DB, and DM do so in zero, three, and four instances, respectively. Notably, FGSM exhibits more successful targeted attacks than state-of-the-art DB and DM, while also achieving a larger average $ME_S$ than DM does. One possible reason for DB could be attributed to its algorithm design, which updates the perturbation based on the absolute steering error, neglecting its sign. Meanwhile, DM is originally designed for online testing within a simulated environment, thereby lacking the ability to generalize effectively to real-world data. Moreover, both DM and DB apply a single, image-agnostic perturbation across the entire video, whereas FGSM and PA generate unique perturbations to each image (image-specific). This may potentially limit DB and DM's efficacy in targeted attacks on models with complex architectures.

\begin{table*}
\caption{Mean Success Rate Result average over all testing videos for three ADSs: CILRS and CILR results are averaged over 7 Carla100 videos. MT results are averaged over 7 real-world driving videos. We tested thresholds $\delta_S$=3.5, 14, 21, 28 in degrees and $\delta_A$=4.6, 13.8, 23.0, 32.2 in km/h.}
\begin{center}
\begin{tabular}{c| c |  cccc | cccc | cccc}
\hline
\multirow{3}{*}{\textbf{Methods}} & & \multicolumn{4}{c|}{\textbf{CILRS}} & \multicolumn{4}{c|}{\textbf{CILR}} & \multicolumn{4}{c}{\textbf{MT}}\\
& $\delta_S $& {$3.5$}& {$14$} & {$21$} & {$28$} & {$3.5$} & {$14$} & {$21$} & {$28$} & {$3.5$} & {$14$} & {$21$} &{$28$}\\
& $\delta_A$  & {$4.6$} & {$13.8$} & 23.0 & 32.2 & $4.6$ & $13.8$ & 23.0 & 32.2 & $4.6$ & $13.8$ & 23.0 & 32.2 \\
\hline
\multirow{2}{*}{\textbf{UniAda}}  & $SR_S$ & \multicolumn{1}{S[table-format=2.2, detect-all]}{\bfseries 96.3*} & \multicolumn{1}{S[table-format=2.2, detect-all]}{\bfseries 89.5*} & \multicolumn{1}{S}{\bfseries 76.5*} & \multicolumn{1}{S|}{\bfseries 45.4*} & \multicolumn{1}{S}{\bfseries 90.5} & \multicolumn{1}{S}{\bfseries 75.3*} & \multicolumn{1}{S}{\bfseries 62.2*} & \multicolumn{1}{S|}{\bfseries 45.1*} & \multicolumn{1}{S}{\bfseries 33.5} & \multicolumn{1}{S}{\bfseries 12.3} & \multicolumn{1}{S[table-format=2.3, detect-all]}{\bfseries 8.34} & 5.34 \\
& $SR_A$ &  \multicolumn{1}{S}{\bfseries 74.7*} & \multicolumn{1}{S}{\bfseries 67.4*} & \multicolumn{1}{S}{\bfseries 62.3*} & \multicolumn{1}{S|}{\bfseries 49.2} & \multicolumn{1}{S}{\bfseries 67.8*} & \multicolumn{1}{S}{\bfseries 61.1*} & \multicolumn{1}{S}{\bfseries 55.7*} & \multicolumn{1}{S|}{\bfseries 43.9*} & \multicolumn{1}{S}{\bfseries 83.7*} & \multicolumn{1}{S}{\bfseries 31.6*} & \multicolumn{1}{S[table-format=2.3, detect-all]}{\bfseries 2.72*} & 0.00\\
\hline
\multirow{2}{*}{\textbf{DM}} & $SR_S$ & 87.6 & 67.2 & 52.5 & 38.9 & 83.4 & 57.3 & 33.5 & 19.6 & 27.9 & 11.2 & 8.32 & 5.34 \\
& $SR_A$ &$-$ & $-$ & $-$ & $-$ & $-$ & $-$ & $-$ & $-$ & $-$ & $-$ & $-$ & $-$\\
\hline
\multirow{2}{*}{DB} & $SR_S$ & 83.3 & 48.8 & 45.6 & 34.3 & 66.1 & 41.7 & 29.2 & 20.2 & 28.2 & 11.2 & 8.32 & 6.02 \\
& $SR_A$ &$-$ & $-$ & $-$ & $-$ & $-$ & $-$ & $-$ & $-$ & $-$ & $-$ & $-$ & $-$\\
\hline
\multirow{2}{*}{PA} & $SR_S$ & 86.2 & 54.6 & 24.3 & 15.1 & 82.0 & 38.6 & 18.6 & 11.0 & 30.7 & 11.9 & 7.68 & 5.38 \\
& $SR_A$ & 67.4 & 60.8 & 52.3 & 41.2 & 55.2 & 48.0 & 41.7 & 28.7 & 40.2 & 9.56  & 0.68 & 0.00\\
\hline
\multirow{2}{*}{FGSM} & $SR_S$ &  3.99& 0.00& 0.00 & 0.00 & 1.32 & 0.00 & 0.00 & 0.00 & 9.89 & 3.67 & 0.79 & 0.00 \\
& $SR_A$ & 43.6 & 35.4 & 28.8 & 18.3 & 37.2 & 27.5 & 21.6 & 11.4 & 16.8 & 0.00 & 0.00 & 0.00 \\
\hline
\multirow{2}{*}{UniEqual} & $SR_S$ & 83.4 & 57.9 & 48.0 & 37.6 & 84.2 & 57.3 & 47.1 & 38.8 & 32.0 & 11.6 & 8.06 & 6.02 \\
& $SR_A$ & 73.7 & 65.9 & 60.1 & 47.6 & 59.4 & 52.6 & 47.3 & 36.3 & 28.3 & 5.85 & 0.00 & 0.00\\
\hline
\end{tabular}
\end{center}
\label{tbl:success rate}
\end{table*}
\textbf{Success Rate Results:} 
The success rate, indicating the percentage of adversarial images successfully exhibiting misleading behavior under a specified threshold, is assessed for different bounds. 
In Table \ref{tbl:success rate}, we select four different bounds for each objective to assess the hierarchy of performance, with $\delta_S$=3.5$^{\circ}$, 14$^{\circ}$, 21$^{\circ}$, 28$^{\circ}$ for the steering angle and $\delta_A$=4.6, 13.8, 23.0, 32.2 km/h for acceleration.

UniAda consistently outperforms four baseline methods across all thresholds and all models, except for the case when $\delta_S$=28$^{\circ}$ under MT model. For example, \tool successfully attack CILRS model to produce a steering prediction error of at least 28$^{\circ}$ for 45.4\% of tested images, compared to DM's 38.9\%, DB's 34.3\%, PA's 15.1\% and FGSM's 0\% at the same threshold. For MT model, \tool can trigger errors on much more images than baselines for acceleration, with at most 43.5\% more images than the second best (i.e., PA). FGSM fails to produce a successful attack under nearly half of the tested thresholds.

\begin{tcolorbox}[right=1px, left=1px, top=1px, bottom=1px]
\textbf{Result 1:} UniAda consistently outperforms all four baseline methods in Mean Error on average for all ADS models. For Success Rate, \tool demonstrates the best performance under 23 out of 24 cases on average. Compared with the state-of-the-art technique, DM, \tool achieves a mean steering error improvement of 6.6$^{\circ}$ in CILRS, 8.6$^{\circ}$ in CILR, and 2.93$^{\circ}$ in MT model.
 \end{tcolorbox}

\begin{table*}[]
\caption{P-value results of Mean Error under CILRS, CILR and MT: results are shown in bold if they have a p-value $\leq 0.05$,  The positive/negative/equal signs indicate that UniAda has a better/worse/equal performance versus UniEqual.}
    \begin{center}
    \begin{tabular}{c|c c|c c|c|c c}
    \hline
    \multirow{2}{*}{\textbf{Videos}} & \multicolumn{2}{c|}{\textbf{CILRS}} & \multicolumn{2}{c|}{\textbf{CILR}} &  \multirow{2}{*}{\textbf{Videos}} & \multicolumn{2}{c}{\textbf{MotionTransformer}} \\
    & $ME_S$ & $ME_A$ & $ME_S$ & $ME_A$ & & $ME_S$ & $ME_A$ \\
    \hline
        Carla Pedestrian & 7.61E-01(+) & 1.67E-01(-) & 8.88E-01(+) & \textbf{3.95E-03(+)} & Dave Curve1 & \textbf{2.58E-08(+)} & \textbf{1.18E-09(+)} \\
        \hline
        Carla White Car & \textbf{3.49E-02(+)} & \textbf{3.19E-02(+)} & 8.05E-01(+) & \textbf{2.64E-02(+)} & Dave Straight1 & \textbf{1.20E-02(+)} & \textbf{1.71E-03(+)} \\
        \hline
        Carla Black Car & \textbf{5.93E-03(+)} & 7.85E-01(-) & \textbf{4.39E-02(+)} & \textbf{1.53E-02(+)} & Udacity Straight1 & \textbf{1.28E-02(+)} & \textbf{1.27E-10(+)} \\
        \hline
        Carla Gray Car & \textbf{6.28E-07(+)} & \textbf{7.48E-07(+)} & \textbf{7.40E-05(+)} & 9.25E-01(+) & Udacity Straight2 & \textbf{2.29E-03(+)} & \textbf{1.43E-05(+)} \\
        \hline
        Carla Blue Car & \textbf{3.58E-02(+)} & 5.38E-01(+) & \textbf{4.99E-05(+)} & \textbf{1.73E-08(+)} & Udacity Curve1 & 6.03E-01(+) & \textbf{4.08E-15(+)} \\
        \hline
        Carla Red Light & \textbf{4.37E-02(+)} & 8.62E-01(+) & \textbf{2.97E-03(+)} & 1.50E-01(-) & Kitti Curve1 & 1.75E-01(-) & \textbf{3.00E-04(+)} \\
        \hline
        Carla Light-blue Car & \textbf{4.28E-07(+)} & 3.81E-01(-) & \textbf{2.63E-04(+)} & \textbf{1.05E-05(+)} & Kitti Straight1 & \textbf{4.75E-04(+)} & \textbf{3.76E-12(+)} \\
        \hline
        Average & \textbf{2.72E-06(+)} & 2.63E-01(+) & \textbf{8.89E-04(+)} & \textbf{1.43E-04(+)} & Average & \textbf{2.71E-05(+)} & \textbf{1.08E-09(+)} \\
        \hline
    \end{tabular}
    \end{center}
    \label{tab:p_values ME}
\end{table*}

\begin{table}[]
    \centering
    \caption{P-value results of Success Rate under CILRS, CILR, and MT. 'NA' denotes unavailable results due to the same performance of UniAda and UniEqual.}
    \begin{tabular}{c|cccc}
    \hline
     $\delta_S$ & 3.5 & 14 & 21 & 28 \\
     $\delta_A$ & 4.6 & 13.8 & 23.0 & 32.2 \\
     \hline
        \multirow{2}{*}{\textbf{CILRS}} & \textbf{2.64E-04(+)} & \textbf{1.39E-04(+)} & \textbf{7.44E-05(+)} & \textbf{1.15E-03(+)} \\
        & \textbf{3.93E-02(+)} & \textbf{4.09E-02(+)} & \textbf{2.32E-02(+)} & 9.92E-02(+) \\
        \hline
         \multirow{2}{*}{\textbf{CILR}} & 5.43E-02(+) & \textbf{5.37E-03(+)} & \textbf{2.98E-02(+)} & \textbf{4.59E-02(+)} \\
         & \textbf{1.37E-03(+)} & \textbf{2.17E-03(+)} & \textbf{1.35E-03(+)} & \textbf{2.29E-02(+)} \\
         \hline
         \multirow{2}{*}{\textbf{MT}} & 3.40E-01(+) & \textbf{9.23E-05(+)} & \textbf{5.56E-09(+)} & 7.32E-02(-) \\
          & \textbf{3.72E-08(+)} & \textbf{7.01E-08(+)} & $\textbf{2.97E-11(+)}$ & NA(=) \\
          \hline
    \end{tabular}
    \label{tab:p_value SR}
\end{table}

\vspace{0.3em}
\textbf{RQ2. Adaptive Weighting Scheme Effectiveness.}
To assess the effectiveness of the Adaptive Weighting Scheme (AWS), we introduce a variant of \tool dubbed UniEqual, which utilizes the same universal multi-objective attack algorithm but employs equal-weighted objectives during the perturbation search process (i.e., UniEqual omits the use of AWS).



Observing the Mean Error results in Table \ref{tbl:me result} and Table \ref{tab:me real-world result}, it is evident that UniAda consistently outperforms UniEqual in both steering and acceleration objectives on average across all ADS models. This indicates that AWS effectively enhances the attack performance in both objectives simultaneously. While in certain videos, UniEqual might achieve a slightly higher attack error in one objective, it becomes apparent that it is due to the lack of focus on the other objective. For instance, in the CILRS model videos ``Black Car'' and ``Light-blue Car'', UniEqual produces 23.9 km/h and 22.0 km/h acceleration error, only 0.1 km/h and 1.0 km/h higher than UniAda, respectively. However, UniAda showcases a steering error of 18.6$^{\circ}$ and 25.9$^{\circ}$, which is significantly higher than UniEqual, with 8.97$^{\circ}$ and 14.7$^{\circ}$, respectively.

For real-world driving videos, \tool consistently outperforms UniEqual in all seven videos, with an average improvement of 9.87 km/h in acceleration error and 0.46$^{\circ}$ in steering error. We can see that AWS maintains the attack strength on steering objective while significantly improves the attack power on acceleration. Moreover, it is noteworthy that in 3 out of 7 cases, UniEqual performs attack in an incorrect direction for the acceleration objective (i.e., misalignment with the specified direction). In contrast, AWS optimizes \tool to successfully attack all tested videos in the correct direction as specified. This is attributed to the AWS feature of balancing the training rate of both objectives by adapting weights, which is to ensure both objectives are thoroughly explored and collectively contribute to the perturbation that significantly affects both objectives simultaneously.

Similar conclusion applies to the Success Rate results. As displayed in Table \ref{tbl:success rate}, \tool outperforms UniEqual in all models across all tested thresholds, except for one instance: $\delta_S$=28$^{\circ}$ under MT model. In this case, perturbation generated by UniEqual can mislead slightly more images than \tool with only 0.68\% improvement. Nevertheless, with the same perturbation, \tool can mislead 55.4\%, 25.8\%, 2.72\% more images than UniEqual at $\delta_A$=4.6 km/h, 13.8 km/h, 23.0 km/h, respectively. 

\textbf{Statistical Significance:} We have conducted two-sample $t$-tests \cite{student1908probable} to assess whether the performance differences between UniAda and UniEqual are statistically significant. Specifically, we treat all five runs of results as a sample group, with the null hypothesis stating that the means of two populations are the same. Each evaluation metric is examined for each driving model for both real-world and simulated driving data. Table \ref{tab:p_values ME} and Table \ref{tab:p_value SR} demonstrate the p-value results for Mean Error and Success Rate, respectively. Individual video results offer a detailed performance comparison for specific scenarios. To examine the statistical significance of overall performance difference under the simulated/real-world driving environment, we also present p-values for the average results (for each run, we compute the average results of 7 videos, and then perform \textit{t}-tests with 5 runs as a group). Results are shown in bold if they have a p-value $\leq$ 0.05, the positive/negative/equal signs indicate that UniAda has a better/worse/equal performance versus UniEqual. For Mean Error p-value results, there are 30 out of 42 cases (exclude average results) \tool statistically significantly outperforms UniEqual. Out of these, 8 cases come from CILRS, 10 cases for CILR, and 12 cases for MT. Moreover, we note that for the videos that UniEqual outperforms \tool, the results are not significant (e.g., ``Black Car'', ``Light-blue Car''). For Success Rate p-value results, in 19 out of 23 cases, \tool is significantly better than UniEqual. 



\begin{figure}
    \centering
    \includegraphics[scale=0.104]{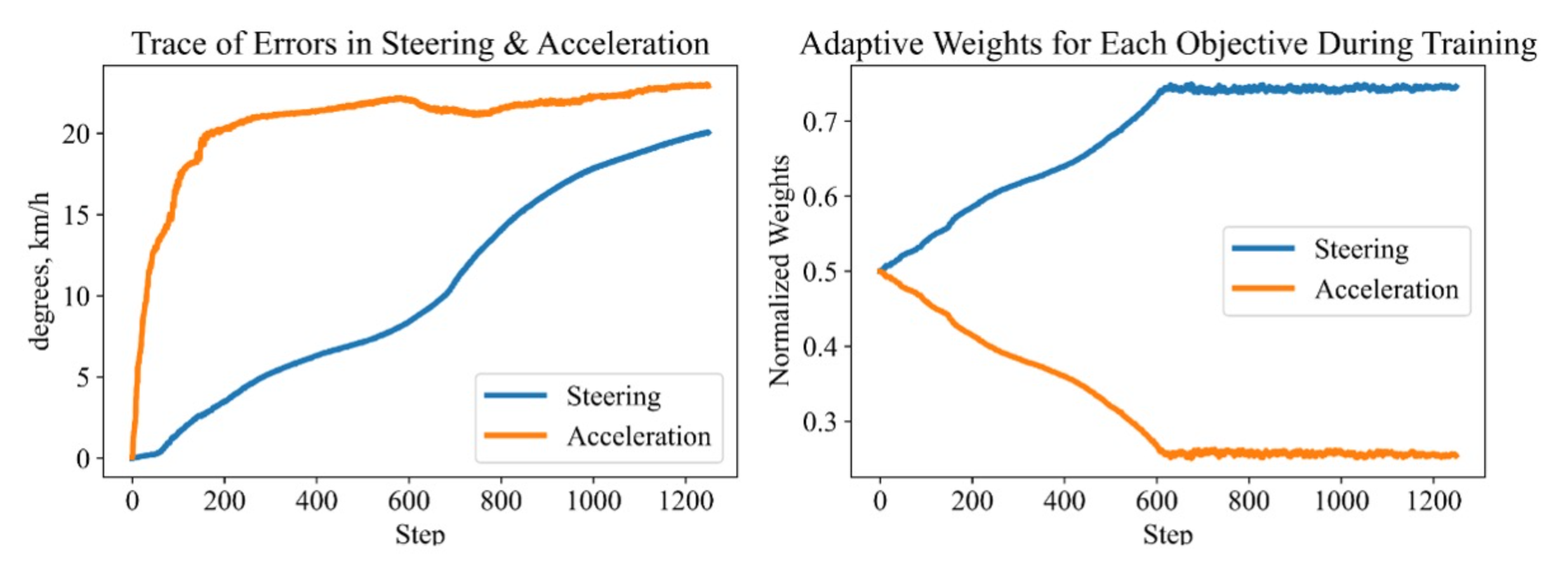}
    \caption{Traces of mean error (left) and AWS normalized weights (right) by \tool attack in CILRS ``Black Car''.}
    \label{fig:AWS}
\end{figure}

To illustrate how AWS functions, we demonstrate an example plot (CILRS ``Black Car'', Fig. \ref{fig:AWS}) of error and weight traces against the perturbation searching process. Each objective weight begins at 0.5 with zero errors. During the initial stages, the attack error in acceleration experiences a substantial improvement, whereas the steering perturbation triggers minimal errors. This discrepancy may be attributed to the higher training rate allocated to acceleration. Consequently, the AWS assigns more weight to the steering objective, driving up the steering error while stabilizing the acceleration error after around 200 steps. The objective weights reach a steady state after approximately 600 steps. At the later stage, although the weights are mainly allocated to steering, the acceleration error still shows a further increase, which implies the perturbation pattern may contain shared information that also triggers errors in acceleration. 


\begin{tcolorbox}[right=1px, left=1px, top=1px, bottom=1px]
\textbf{Result 2:} AWS assists UniAda to improve the performance in both objectives simultaneously in most cases. On average, AWS improves $ME_S$ by 7.0$^{\circ}$, 3.0$^{\circ}$, 0.46$^{\circ}$, and $ME_A$ by 0.4 km/h, 2.2 km/h, 9.87 km/h, for CILRS, CILR, and MT, respectively.
 \end{tcolorbox}

\vspace{0.3em}
\textbf{RQ3. Multi-objective Attack Effectiveness.}
To assess the effectiveness of the multi-objective attack, we conduct experiments involving two single-objective attacks and compare their results with UniAda for CILRS and MT models. The single-objective attacks utilize the same algorithm as \tool, differing solely in the loss function employed. We report the average results over the corresponding testing videos, summarized in Table \ref{tbl:ablation study}, with attack directions set at $d_S$=1 and $d_A$=1.

\begin{table}[h]
\caption{Ablation study on CILRS and MT models, averaged over testing videos}
\begin{center}
\begin{tabular}{c|ccc|ccc}
\toprule
& & CILRS & & & MT & \\
 & $L_{Steer}$ & $L_{Acc}$ & UniAda & $L_{Steer}$ & $L_{Acc}$ & UniAda\\
 \hline
  $ME_S$ & 23.7 & - &  29.2 & 3.26 & - & 3.54 \\
  $ME_A$ & - & 19.5 &  23.0 & - & 9.82 & 11.0 \\
 \bottomrule
\end{tabular}
\end{center}
\label{tbl:ablation study}
\end{table}




On average, UniAda consistently outperforms both single-objective attacks. Specifically, in terms of steering, \tool induces a steering angle misdirection of 5.5$^{\circ}$ more than the single-objective attack focused on steering ($L_{Steer}$). Similarly, for acceleration, \tool generates an acceleration error around 3.5 km/h higher compared to the single-objective acceleration attack ($L_{Acc}$). Similar results for MT model, where single-objective attack is 0.28$^{\circ}$ and 1.18 km/h less than that of \tool. This outcome suggests that the perturbation patterns sought for steering and acceleration objectives complement each other, showcasing the advantage of multi-objective attacks.

\begin{tcolorbox}[right=1px, left=1px, top=1px, bottom=1px]
\textbf{Result 3:} Compared to single objective attacks, the multi-objective attack boosts the performance for both objectives simultaneously for both simulated and real-world videos.
 \end{tcolorbox}



\section{Discussion}
\subsection{Transferability of Adversarial Attacks}
\begin{figure}
    \centering
    \includegraphics[scale=0.16]{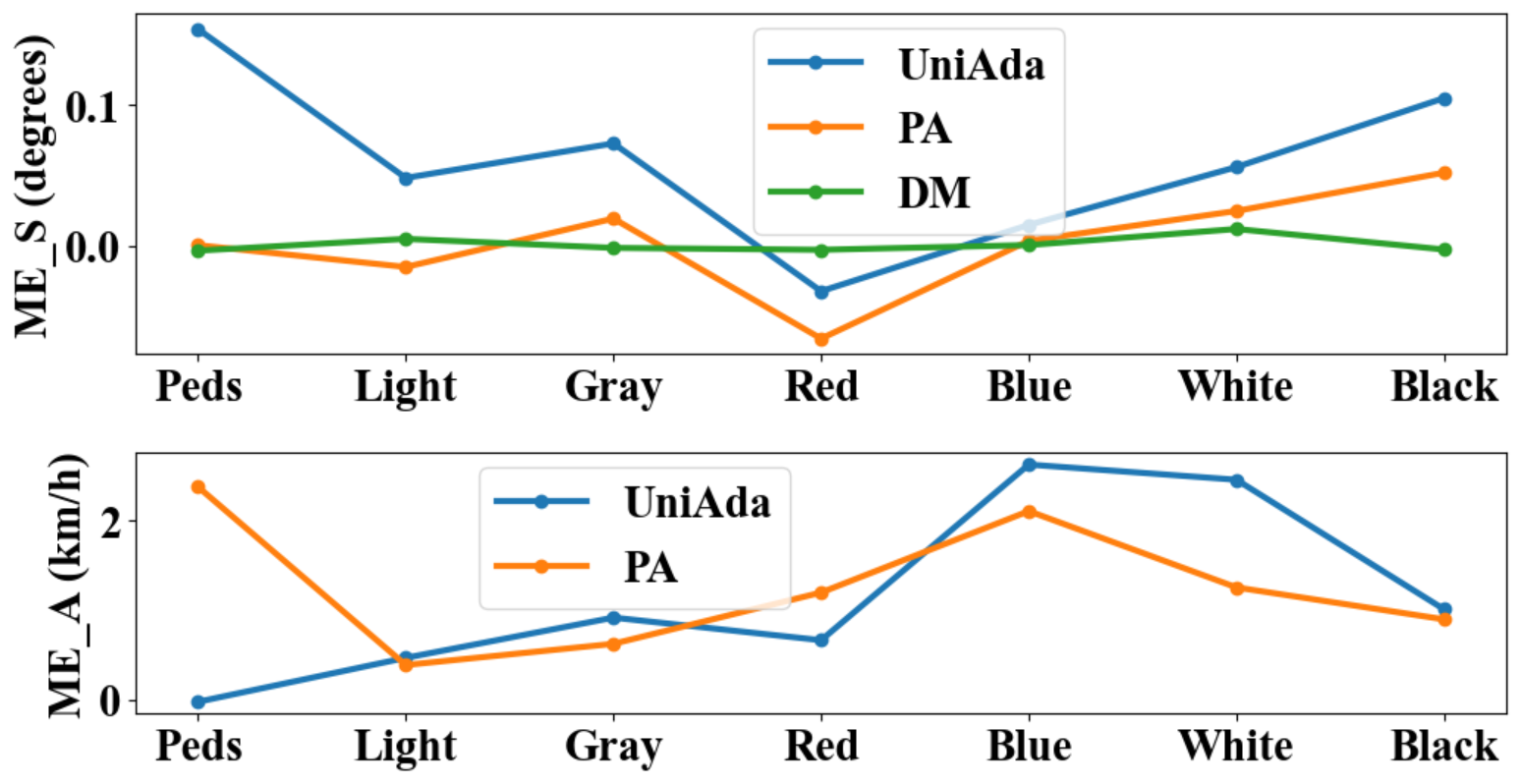}
    \caption{Transferability of Adversarial attacks: Mean Error results in Steering (top) and Acceleration (bottom) of \tool, DM, PA. Note that DM results for $ME_A$ is not available.}
    \label{fig:transfer}
\end{figure}

In this section, we analyze the transferability of the generated adversarial perturbation on different autonomous driving models. Specifically, we explore the attack effectiveness of the perturbation generated with CILRS on CILR model for each video in Carla100 dataset. We compare the Mean Error results of UniAda with its two closest competitors, DM and PA, shown in Figure \ref{fig:transfer}.

As shown in the plot, for $ME_S$, the perturbation generated by \tool with CILRS model under test can cause the highest error when attack CILR model in most cases. Only in the video "Red Light", DM produces a better performance. For $ME_A$, \tool outperforms PA in 5 out of 7 cases. Only in "Pedestrian", PA outperforms \tool by a relatively large margin.  However, we can see that all three techniques exhibit weak transferability, a phenomenon frequently observed under the white-box setting owing to the attack's specificity to the internal structure of the model \cite{wang2021enhancing}. Possible directions to address this issue include using advanced gradient calculation \cite{dong2018boosting}, or applying various input transformations \cite{xie2019improving}. We leave the improvement of transferable attacks for further work. 


\subsection{Limitations}
\tool is a gradient-based approach that requires white-box access, leading to less generalizability on DNNs with restricted access or knowledge. Moreover, our findings on real-world videos is limited by one victim DNN, MotionTransformer. The choice of MotionTransformer is due to its complex architecture (i.e., incorporates RGB and optical flow images to learn both position and motion information) and its competitive performance (e.g., surpassing CNN-based DAVE2 model). Our findings are also limited to the parameter space we explored, capturing key factors influencing the techniques' performance but is still limited in scope. 

\subsection{Threats to Validity}
\textbf{Internal Validity:} The threats to internal validity can be attributed to the implementation quality of baselines and reproducibility of our method. To mitigate this, we open-source our implementation to facilitate checking and reproducibility. 

\textbf{External Validity:} The first threat to external validity to our experimental conclusions is our selected datasets (i.e., 7 simulated videos from Carla100 dataset, 7 real-world videos from Dave, Kitti, and Udacity datasets) and DNN models (i.e., CILRS, CILR, MT). We tried to alleviate this threat as follows: (1) the selected datasets contain both simulated and real-world driving scenarios. They are also widely used in the previous research \cite{deepbillboard, physgan, exploring}; (2) the three autonomous driving models have achieved competitive performance in the field. They are constructed with different architectures and different numbers of layers. For example, CILR and CILRS use ResNet only to learn RGB data information, while MT is composed of more complex architecture that incorporates optical flow with RGB images to capture both position and motion information. Therefore, our experimental conclusions should generally hold with other driving datasets and models.

The second external validity concern is the selection of baselines. To mitigate this, we compare \tool with DeepManeuver \cite{von2023deepmaneuver}, Perturbation Attack \cite{zhang2021evaluating}, DeepBillboard \cite{deepbillboard}, FGSM \cite{fgsm} that are either state-of-the-art (e.g., DeepManeuver, published in 2023) or representative (e.g., DeepBillboard, FGSM) in this field.

\textbf{Construct Validity:} Construct validity concerns whether the chosen evaluation metrics accurately capture the intended effect. In our study, our goal is to design an attack technique that generates image-agnostic perturbations capable of inducing errors in the model under test. We mitigate the threat by examining the attack effectiveness on two metrics,  mean error and success rate, which are widely used in the literature \cite{wu2023adversarial, deng2020analysis, physgan, deepbillboard}. The mean error can reflect the average attack strength, however, it might be biased by some outliers and cannot represent the performance of most input images. To counter this threat, we also measure the success rate, which manifests the universality of the attack (i.e., whether the generated perturbation can affect most input images) by examining the percentage of images that are successfully attacked under different thresholds.

\section{Conclusion}
In this paper, we examine the reliability and security problems raised by adversarial attacks on three End-to-End autonomous driving systems. Our proposed method, \tool, conducts novel and comprehensive testing by addressing three main limitations from existing literature: limited vehicle controls testing, constrained testing scenarios, and image-specific or unnatural perturbations. We alleviate them accordingly by (1) performing testing on steering and acceleration controls simultaneously through the multi-objective attack with Adaptive Weighting Scheme, (2) examining both simulated and real-world driving scenarios in urban traffic, and (3) generating image-agnostic (i.e., universal) and human-eye-imperceptible perturbation through joint optimization. The effectiveness of \tool is compared with four baselines in mean error and success rate metrics. Compared with its closest competitor, DeepManeuver, UniAda achieves an improvement of 6.6$^{\circ}$, 8.6$^{\circ}$, and 2.9$^{\circ}$ in steering error on CILRS, CILR, and MT ADSs, respectively.

\bibliographystyle{ieeetr}

\bibliography{ref}{}



\end{document}